\documentclass[aps,prd,preprintnumbers,superscriptaddress,nofootinbib]{revtex4-2} 
\usepackage[utf8]{inputenc}
\usepackage{amssymb,amsmath,amsfonts}
\usepackage{multirow}
\usepackage{slashed} 
\usepackage{graphicx}
\usepackage{subfigure}
\usepackage{hyperref}
\hypersetup{colorlinks=true,linkcolor=purple,anchorcolor=blue,citecolor=blue, filecolor=blue,urlcolor=red,bookmarksnumbered=true,
	pdfview=FitB
}
\usepackage{multirow}
\usepackage{color}
\usepackage{xcolor}
\usepackage{ulem}
\usepackage{csquotes}
\colorlet{purple1}{blue!70!red}
\colorlet{darkred}{red!50!black}
\usepackage{float}

\def\orcid#1{\kern .08em\href{https://orcid.org/#1}{\includegraphics[keepaspectratio,width=0.7em]{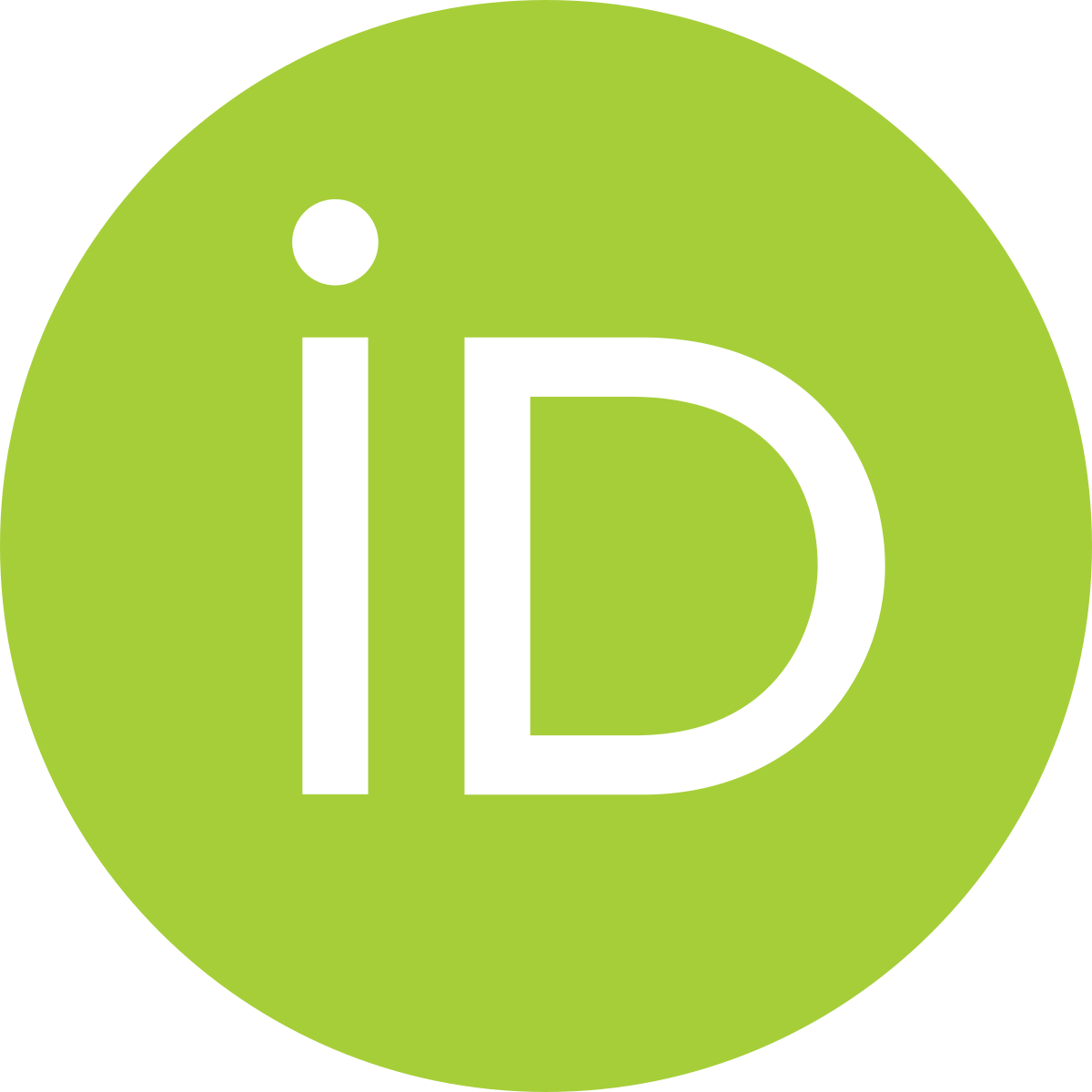}}}

\def\orcid#1{\kern .08em\href{https://orcid.org/#1}{\includegraphics[keepaspectratio,width=0.7em]{ORCID_iD.png}}}

\begin{document}
  	\title{$\phi$-meson spectroscopy and diffractive production using two Schr\"odinger like equations on the light-front}%

		\author{Bheemsehan~Gurjar\orcid{0000-0001-7388-3455}}
		\email{gbheem@iitk.ac.in} 
		\affiliation{Department of Physics, Indian Institute of Technology Kanpur, Kanpur-208016, India}
     	\author{Chandan~Mondal\orcid{0000-0002-0000-5317}}
		\email{mondal@impcas.ac.cn}
		\affiliation{Institute of Modern Physics, Chinese Academy of Sciences, Lanzhou 730000, China}
		\affiliation{School of Nuclear Science and Technology, University of Chinese Academy of Sciences, Beijing 100049, China}
		
		\author{Satvir~Kaur\orcid{0000-0002-7643-5970}}
		\email{satvir@impcas.ac.cn}
		\affiliation{Institute of Modern Physics, Chinese Academy of Sciences, Lanzhou 730000, China}
		\affiliation{School of Nuclear Science and Technology, University of Chinese Academy of Sciences, Beijing 100049, China}
\begin{abstract}
We show that the holographic Schr\"odinger equation of light-front chiral QCD, together with the 't Hooft equation of (1+1)-dimensional QCD in the large $N_c$ limit, can simultaneously describe the $\phi$-meson mass spectroscopy as well as diffractive cross-section. We compute the $\phi$-meson diffractive cross-section by  utilizing its resulting light-front wave functions (LFWFs), in conjunction with the color glass condensate (CGC) dipole scattering amplitude. Our predictions for the diffractive cross sections show good agreement with the existing experimental data from HERA at various energies from H1 and ZEUS collaborations. Additionally, we show that the obtained $\phi$-meson LFWFs effectively describe its various properties, including the decay constant, distribution amplitudes, electromagnetic form factors, charge radius, and magnetic and quadrupole moments. 
\end{abstract}	
\maketitle
%
%%%%%%%%%%%%%%%%%
\section{Introduction}
%%%%%%%%%%%%%%%%%
Diffractive vector meson production in exclusive processes serves as a powerful probe of gluon saturation in the small-$x$ region~\cite{Jalilian-Marian:1996mkd,Ryskin:1992ui}. This phenomenon has been extensively studied within the framework of the Color Glass Condensate (CGC) effective field theory~\cite{Gelis:2010nm,Iancu:2002xk,McLerran:2001sr}. Saturation-based models have demonstrated remarkable success in describing high-precision data from the Hadron–Electron Ring Accelerator (HERA)~\cite{Golec-Biernat:1998zce,Levin:2000mv} and are pivotal for predictions at upcoming facilities such as the Large Hadron Electron Collider (LHeC)~\cite{LHeCStudyGroup:2012zhm}, the Electron–Ion Collider (EIC) in the United States~\cite{Aschenauer:2014twa,Accardi:2012qut,AbdulKhalek:2021gbh}, and the Electron–Ion Collider in China (EIcC)~\cite{Anderle:2021wcy}.

The dipole model provides a unified theoretical framework to investigate both inclusive and exclusive diffractive processes in electron-proton scattering at HERA~\cite{Mueller:1993rr,Nemchik:1996cw}. In this approach, the virtual photon splits into a quark-antiquark pair, forming a dipole~\cite{PhysRevD.74.074016}, which interacts with the proton via gluon exchange. This interaction may lead to the reformation of the dipole into a real photon or a vector meson in the final state~\cite{PhysRevD.78.014016}. The diffractive cross-section for vector meson production can be factorized into the product of the light-front wavefunctions (LFWFs) of the incoming virtual photon and the outgoing vector meson, and the dipole-proton scattering amplitude~\cite{PhysRevD.74.074016}. The latter encodes the QCD dynamics of the dipole-proton interaction and is typically derived by solving the Balitsky-Kovchegov (BK) equation~\cite{Balitsky:1995ub}.
While the virtual photon wavefunction is well-established in perturbative QCD~\cite{PhysRevD.22.2157}, obtaining the vector meson wavefunction as a nonperturbative object remains an open question. 
In this work, we determine the vector meson wavefunctions by solving the holographic light-front Schrödinger equation~\cite{Brodsky:2014yha}, supplemented with longitudinal dynamics from the 't Hooft equation in $(1+1)$-dimensional QCD at large $N_c$~\cite{tHooft:1974pnl}.

Holographic light-front QCD (hLFQCD) is formulated within the chiral limit of light-front QCD, establishing a correspondence between strongly coupled (3+1)-dimensional light-front QCD and weakly interacting string modes in (4+1)-dimensional anti-de Sitter (AdS) space. For a detailed review of hLFQCD, see Ref.~\cite{Brodsky:2014yha}.
A key prediction of hLFQCD is that the pion, as the lightest bound state, is massless in the chiral limit. Additionally, meson masses are predicted to follow universal Regge trajectories, consistent with experimental data. The slopes of these trajectories are determined by the strength of the confining potential, represented by the parameter $\kappa$. The confining potential in physical spacetime is derived from a dilaton field that breaks the conformal symmetry of AdS space. A quadratic dilaton in the fifth dimension of AdS space corresponds to a light-front harmonic oscillator in physical spacetime, providing a phenomenologically successful framework.
The mass scale parameter $\kappa$ is determined by fitting the experimentally observed slopes of Regge trajectories for various meson groups, with a consistent value of $\kappa \simeq 0.5$ GeV for all light mesons~\cite{Brodsky:2014yha}.

To extend beyond the semiclassical approximation, Brodsky and de Téramond introduced the invariant mass ansatz (IMA) to incorporate nonzero quark masses~\cite{Brodsky:2008pg}. Using the IMA, the mass shifts of mesons can be calculated as first-order perturbations. Notably, the predicted mass shifts for the pion and kaon align with their physical masses. These results were obtained by setting the scale parameter to $\kappa = 0.54$ GeV, with light quark masses of $m_{u/d} = 0.046$ GeV and $m_{s} = 0.357$ GeV, which approach zero in the chiral limit~\cite{BRODSKY20151}.

Previous studies~\cite{PhysRevLett.109.081601,PhysRevD.94.074018} have made predictions for vector meson production by combining the holographic wave function with the IMA and the CGC dipole cross section~\cite{PhysRevD.78.014016}.
Reference~\cite{PhysRevLett.109.081601} examined $\rho$-meson production using a light quark mass of $m_{q} = 0.14$ GeV, which aligns with the parameters fitted to the CGC dipole cross section~\cite{PhysRevD.74.074016,Jeffrey} from inclusive deep inelastic scattering (DIS) data~\cite{ZEUS:2001mhd,H1:2000muc}. The most recent dipole cross section analyses have used the 2010 DIS data from HERA~\cite{H1:2009pze}. In Ref.~\cite{PhysRevD.78.014016}, it was acknowledged that the DIS data prefers lower light quark masses, but the effective quark mass of $m_{q} = 0.14$ GeV provides a good fit to the 2001 DIS structure function data. A more recent study~\cite{Contreras:2015joa} demonstrated that both the current quark mass and the effective mass of $m_{q} = 0.14$ GeV accurately fit the 2010 DIS structure function data~\cite{H1:2009pze}.
In Ref.~\cite{PhysRevD.94.074018}, the authors revisited the CGC dipole model and fitted it to the 2015 HERA inclusive DIS data, incorporating light quark masses. They studied cross sections for diffractive $\rho$ and $\phi$ meson production using the fitted dipole cross section~\cite{IANCU2004199}, perturbatively calculated photon LFWFs~\cite{PhysRevD.22.2157,PhysRevD.55.2602}, and the holographic meson LFWFs with IMA, which do not include dynamical information of the meson in the longitudinal direction~\cite{PhysRevLett.102.081601}.

In this work, we incorporate nonzero light quark masses by employing chiral symmetry breaking via the 't Hooft equation, accounting for longitudinal modes in $(1+1)$-dimensional QCD under the large $N_c$ approximation. By combining the transverse modes derived from the light-front Schrödinger equation with the longitudinal modes of the 't Hooft equation, it is possible to reconstruct spherically symmetric LFWFs~\cite{Ahmady:2021lsh,Ahmady:2021yzh,Ahmady:2022dfv,Gurjar:2024wpq}. 
The combined holographic Schrödinger equation and the 't Hooft equation offer a reliable framework for describing the $\phi$-meson spectrum using the universal parameter $\kappa$~\cite{Brodsky:2014yha}. We demonstrate that this approach can simultaneously account for various properties of the $\phi$-meson, such as its decay constant, parton distribution amplitude (PDA), electromagnetic form factors (EMFFs) and radii. Furthermore, in conjunction
with the CGC dipole cross-section, the resulting wave functions provide a good description of the HERA data on diffractive $\phi$-meson electroproduction.

The rest of the paper is organized as follows: Sec.~\ref{sec:dipole model} provides a brief overview of the color dipole model. In Sec.~\ref{sec:hLFQCD}, we describe the computation of meson LFWFs using holographic light-front QCD, incorporating longitudinal dynamics via the 't Hooft equation. Section~\ref{sec:results} presents the results for $\phi$-meson mass spectroscopy, diffractive cross-section, distribution amplitudes, and electromagnetic form factors. Finally, we conclude the paper in Sec.~\ref{sec:conclusion}.

%%%%%%%%%%%%%%%%%
\section{THE DIPOLE MODEL of EXCLUSIVE VECTOR MESON PRODUCTION}\label{sec:dipole model}
%%%%%%%%%%%%%%%%%
Within the dipole model framework, the scattering amplitude for exclusive vector meson production can be factorized into the LFWFs of the incoming virtual photon and the produced vector meson, combined with the dipole cross-section~\cite{PhysRevD.74.074016},
\begin{eqnarray}
\Im \mbox{m}\, \mathcal{A}^{\gamma^{\ast}p\rightarrow Vp}_\Lambda(s,t;Q^2)  
 &=&   \sum_{h, \bar{h}} \int {\mathrm d}^2 {\mathbf r}_{\perp} \; {\mathrm d} x \; \Psi^{\gamma^*,\Lambda}_{h, \bar{h}}(x,{\mathbf r}_{\perp};Q^2)  \Psi^{V,\Lambda}_{h, \bar{h}}(x,{\mathbf r}_{\perp})^* e^{-i x \mathbf{r}_{\perp} \cdot \mathbf{\Delta}} \mathcal{N}(x_{\text{m}},\mathbf{r}_{\perp}, \mathbf{\Delta}),
\label{eq:amplitude-VMP} 
\end{eqnarray}
where, $\Lambda=0,\pm 1$ denotes the longitudinal ($\Lambda=0$) or transverse ($\Lambda=\pm 1$) polarizations of the vector meson or photon (with virtuality $Q^{2}$), $t=\Delta^{2}$ represents the square of momentum transfer, and $h(\bar{h})$ denotes the helicities of the quark (antiquark). The variable $\mathbf{r}_{\perp}$ refers to the transverse size of the dipole, $x$ defines the longitudinal momentum fraction of the quark, $\mathcal{N}$ stands for the dipole scattering amplitude, and $x_{\rm m}$ denotes the modified Bjorken variable~\cite{PhysRevD.88.074016}. $\Psi^{\gamma\ast,\Lambda}_{h,\bar{h}}$ and $\Psi^{V,\Lambda}_{h,\bar{h}}$ represent the LFWFs of the virtual photon and the produced vector meson, respectively. The differential cross-section can be expressed in terms of the scattering amplitude~\cite{PhysRevD.78.014016,ZEUS:2007iet}:
\begin{eqnarray}\label{eq:diffcrosssect} 
	\frac{{\rm d}\sigma_{\Lambda}^{\gamma^{\ast}p\rightarrow Vp}}{{\rm d}t}= \frac{1}{16\pi} 
	[\Im\mathrm{m} \mathcal{A}^{\gamma^{\ast}p\rightarrow Vp}_\Lambda(s, t=0)]^2 \; (1 + \beta_\Lambda^2) \exp(-B_D t),
\end{eqnarray}
where $\beta_{\Lambda}$ is the ratio of the real to imaginary parts of the scattering amplitude. This parameter accounts for phenomenological corrections to match the experimental data and is given by~\cite{PhysRevD.62.014022,PhysRevD.78.014016,PhysRevD.87.034002}:
\begin{eqnarray}
\beta_{\Lambda}=\tan\left(\frac{\pi}{2}\alpha_{\Lambda}\right),\,~~~~~~\textrm{with}~~~~~~\alpha_{\Lambda}\equiv \frac{\partial \textrm{log}|\Im\mathrm{m} \mathcal{A}_{\Lambda}^{\gamma^{\ast}p\rightarrow Vp}|}{\partial \textrm{log}(1/x_{\rm m})},
\end{eqnarray}
and $B_{D}$ represents the diffractive slope parameter. This parameter characterizes the $t$-dependence of the differential cross-section and is consistent with experimental data. The slope parameter $B_{D}$ is expressed as~\cite{PhysRevD.94.074018,PhysRevD.78.014016}:
\begin{eqnarray}
B_D = N\left[ 14.0 \left(\frac{1~\mathrm{GeV}^2}{Q^2 + M_{V}^2}\right)^{0.2}+1\right],
\label{eq:Bslope}
\end{eqnarray}
with $N=0.55~\mathrm{GeV}^{-2}$. This parameterization aligns with ZEUS data on vector meson production~\cite{ZEUS:2007iet,PhysRevD.87.034002}, and we adopt the same form for $\phi$-meson production.

Numerous parameterizations of the dipole cross-section exist in the literature, derived from various theoretical considerations and often influenced by the Golec-Biernat–Wüsthoff (GBW) model~\cite{PhysRevD.60.114023}. Among these, an earlier proposal for the dipole cross-section was introduced in Ref.~\cite{IANCU2004199}, commonly referred to as the CGC dipole model. This model interpolates between the solutions of the Balitsky–Fadin–Kuraev–Lipatov (BFKL) equation, applicable to small dipole sizes, and the Levin–Tuchin solution (see Ref.~\cite{Lipatov:1976zz}) of the BK equation~\cite{PhysRevD.60.034008}, which is relevant deep within the saturation region for larger dipoles. Meanwhile, the impact parameter ($b_{\perp}$) integrated CGC dipole cross section is expressed as~\cite{PhysRevD.78.014016}:
\begin{eqnarray}
	\hat{\sigma}(x_{\text{m}},r_{\perp}) = \sigma_0 \, { {\mathcal{N}} (x_{\text{m}}, r_{\perp} Q_s) }
	\label{eq:sigmaCGC}
\end{eqnarray}
with
\begin{eqnarray}
	{\mathcal{N}} (x_{\text{m}},r_{\perp} Q_s) &=& \begin{cases}
 { \mathcal N}_0 \left ( \frac{{ r_{\perp} Q_s}}{2}\right)^ {2 \left [ \gamma_s + \frac{{\mathrm ln}  (2 / r_{\perp} Q_s)}{\kappa_0 \, \lambda \, {\mathrm ln} (1/x_{\text{m}})}\right]}  ~~~~~~~{\rm for }~~~~~~~r_{\perp} Q_s \leq 2 \,,\\
{ 1- \exp[-{\mathcal A} \,{\mathrm ln}^2 ( {\mathcal B} \, r_{\perp} Q_s)]}~~~~~~~~~~~{\rm for }~~~~~~~r_{\perp} Q_s > 2\,,
	\end{cases}
	\label{eq:dipolcross}
\end{eqnarray}
where, $Q_{s}=(x_{0}/x_{\rm m})^{\lambda/2}$ GeV is the saturation scale, The coefficients $\mathcal{A}$ and $\mathcal{B}$ in Eq.~\eqref{eq:dipolcross} are determined from the condition that ${\mathcal{N}}(x_{\rm m},r_{\perp}Q_{s})$ and its derivative with respect to $r_{\perp}Q_s$ are continuous at $r_{\perp}Q_s=2$. This results in
\begin{eqnarray} \label{eq:AandB}
 \mathcal{A} = -\frac{\mathcal{N}_0^2\gamma_s^2}{(1-\mathcal{N}_0)^2\ln(1-\mathcal{N}_0)}, \qquad \mathcal{B} = \frac{1}{2}\left(1-\mathcal{N}_0\right)^{-\frac{(1-\mathcal{N}_0)}{\mathcal{N}_0\gamma_s}}.
\end{eqnarray}
The parameters $\mathcal{N}_{0}$ and $\kappa_0$ are fixed at $0.7$ and $9.9$, respectively, based on the LO BFKL analysis~\cite{Iancu:2003ge}. The remaining parameters, $\sigma_0$, $\lambda$, $x_0$, and $\gamma_s$, were fitted to the proton $F_2$ structure function data for $x_{\rm Bj} \leq 0.01$ and $0.045 \leq Q^2 \leq 45~{\rm GeV}^2$ at HERA~\cite{ZEUS:2000sac,ZEUS:2001mhd,H1:2015ubc}. In this work, we adopt the same parameter values as those determined in Ref.~\cite{PhysRevD.94.074018}.

To calculate the scattering amplitude for the exclusive production of the $\phi$-meson, as outlined in Eq.~(\ref{eq:amplitude-VMP}), it is essential to employ the LFWFs of the virtual photon. These photon LFWFs can be derived perturbatively using light-front QED~\cite{PhysRevD.22.2157}, and are expressed as follows:%,PhysRevD.55.2602,Kulzinger:1998hw},
\begin{eqnarray}\label{eq:photonLFWFs}\nonumber
\Psi^{\gamma,\Lambda=0}_{h,\bar{h}}(x,{\mathbf r}_{\perp}; Q^2, m_q)  &=& \sqrt{\frac{N_{c}}{4\pi}}\delta_{h,-\bar{h}}e\, e_{q}2 x(1-x) Q \frac{K_{0}(\epsilon {\mathbf r}_{\perp})}{2\pi}, 
\label{photonwfL} \\
\Psi^{\gamma,\Lambda=\pm 1}_{h,\bar{h}}(x,{\mathbf r}_{\perp}; Q^2, m_q) &=& \pm \sqrt{\frac{N_{c}}{2\pi}} e \, e_{q} 
 \big[i e^{ \pm i\theta_{{\mathbf r}_{\perp}}} (x \delta_{h\pm,\bar{h}\mp} -  (1-x) \delta_{h\mp,\bar{h}\pm}) \partial_{{\mathbf r}_{\perp}}   +  m_{q} \delta_{h\pm,\bar{h}\pm} \big]\frac{K_{0}(\epsilon {\mathbf r}_{\perp})}{2\pi},
\end{eqnarray}
where $\epsilon=\sqrt{x(1-x)Q^{2} + m_{q}^{2}}$ and $e=\sqrt{4\pi\alpha_{\rm em}}$, with the fine structure constant $\alpha_{\rm em}$; $e_{q}$ and $m_q$ represent the effective charge and mass of the quark, respectively. $N_{c}$ represents the color factor, while $K_{0}$ denotes the Bessel function of the second kind. The total cross-section is expressed as a linear combination of the transverse and longitudinal components, determined by integrating each component (as given in Eq.~\eqref{eq:diffcrosssect}) over $t$:
\begin{eqnarray}
\sigma_{\textrm{tot}}^{\gamma^{\ast}p\rightarrow Vp}(x,Q^{2})=\sigma_{\Lambda=\pm 1}^{\gamma^{\ast}p\rightarrow Vp}(x,Q^{2})+\mathcal{\varepsilon} \sigma_{\Lambda=0}^{\gamma^{\ast}p\rightarrow Vp}(x,Q^{2}),
\end{eqnarray}
where $\varepsilon$ represents the photon polarization parameter, which has an average value of $\langle\varepsilon\rangle=0.98$ in the kinematic range of the HERA experiment for $\phi$-meson production~\cite{H1:2009cml}. We use the same value of $\varepsilon$ to predict the total cross-section and compare our results with the HERA data.

%====================
\section{Holographic meson wave functions and mass spectroscopy}
\label{sec:hLFQCD}
%====================
The explicit form of hadronic LFWFs cannot be derived using perturbation theory. Various approaches for modeling the nonperturbative wave functions of mesons have been proposed in the literature~\cite{Brodsky:2014yha,PhysRevD.55.2602,Nemchik:1996cw,PhysRevLett.109.081601}. Among these, the boosted Gaussian wave function~\cite{Nemchik:1996cw,PhysRevD.69.094013} is one of the most widely used and has been successfully employed in recent studies to effectively reproduce cross-section data for diffractive $\rho$, $\phi$, and $J/\Psi$ production~\cite{PhysRevD.69.074016,PhysRevD.87.034002}. Following a similar approach to the photon LFWF, the spin-improved LFWFs for longitudinally ($\Lambda=0$) and transversely polarized ($\Lambda=\pm 1$) vector mesons are expressed as~\cite{PhysRevD.69.094013,PhysRevLett.109.081601,Kaur:2020emh,PhysRevD.102.034021},
\begin{eqnarray}
\Psi^{V,\Lambda=0}_{h,\bar{h}}(x,{\mathbf r}_{\perp}) =  \frac{1}{2} \delta_{h,-\bar{h}}  \bigg[ 1 + \frac{m_{q}^{2} -\nabla_{{\mathbf r}_{\perp}}^{2}}{x(1-x)M^2_{V}} \bigg] \Psi(x,{\mathbf r}_{\perp}),
\label{eq:LFWFlong}
\end{eqnarray}
and
\begin{eqnarray}
\Psi^{V, \Lambda=\pm 1}_{h,\bar{h}}(x,{\mathbf r}_{\perp}) = \pm \bigg[  i e^{\pm i\theta_{{\mathbf r}_{\perp}}}  ( x \delta_{h\pm,\bar{h}\mp} - (1-x)  \delta_{h\mp,\bar{h}\pm})  \partial_{{\mathbf r}_{\perp}}+ m_{q}\delta_{h\pm,\bar{h}\pm} \bigg] \frac{\Psi(x,{\mathbf r}_{\perp})}{2 x (1-x)}, 
\label{eq:LFWFtrans}
\end{eqnarray}
respectively, where $M_{V}$ is the mass of the vector meson, $\Psi$ is the spin independent part of the  wave functions. 
%%%%%%%%%%%%%%
\begin{table}
\caption{Quantum numbers and mass spectra of the $\phi$-meson family {{with $S=1$}}. $M_\perp$ and $M_\parallel$ are obtained by solving the transverse and longitudinal dynamical equations,  the holographic Schr\"odinger equation and the 't Hooft equation, respectively.}
\vspace{0.15cm}
\centering
\begin{tabular}{|c c c c c c c c|}
\hline
$J^{P(C)}$ & Name & $n_\perp$ & $n_\parallel$ & $L$ & $M_\parallel$ [MeV] & $M_\perp$ [MeV] & $M_{\rm tot}$ [MeV] (This work)\\
% &  &  &  &  &   &  &  \\
\hline
$1^{--}$ & $\phi(1020)$ & 0 & 0 & 0 & 752 & 740 & 1054 \\
$3^{--}$ & $\phi_3(1850)$ & 0 & 4 & 2 & 921 & 1654 & 1893 \\
$1^{--}$ & $\phi(1680)$ & 1 & 2 & 0 & 852 & 1281 & 1538\\
$1^{--}$ & $\phi(2170)$ & 3 & 6 & 0 & 979 & 1957 & 2188 \\
\hline
\end{tabular}
\label{Spectroscopy}
\end{table}
%%%%%%%%%%%%%%
%%%%%%%%%%%%%
\begin{figure}
    \centering
    \includegraphics[scale=0.5]{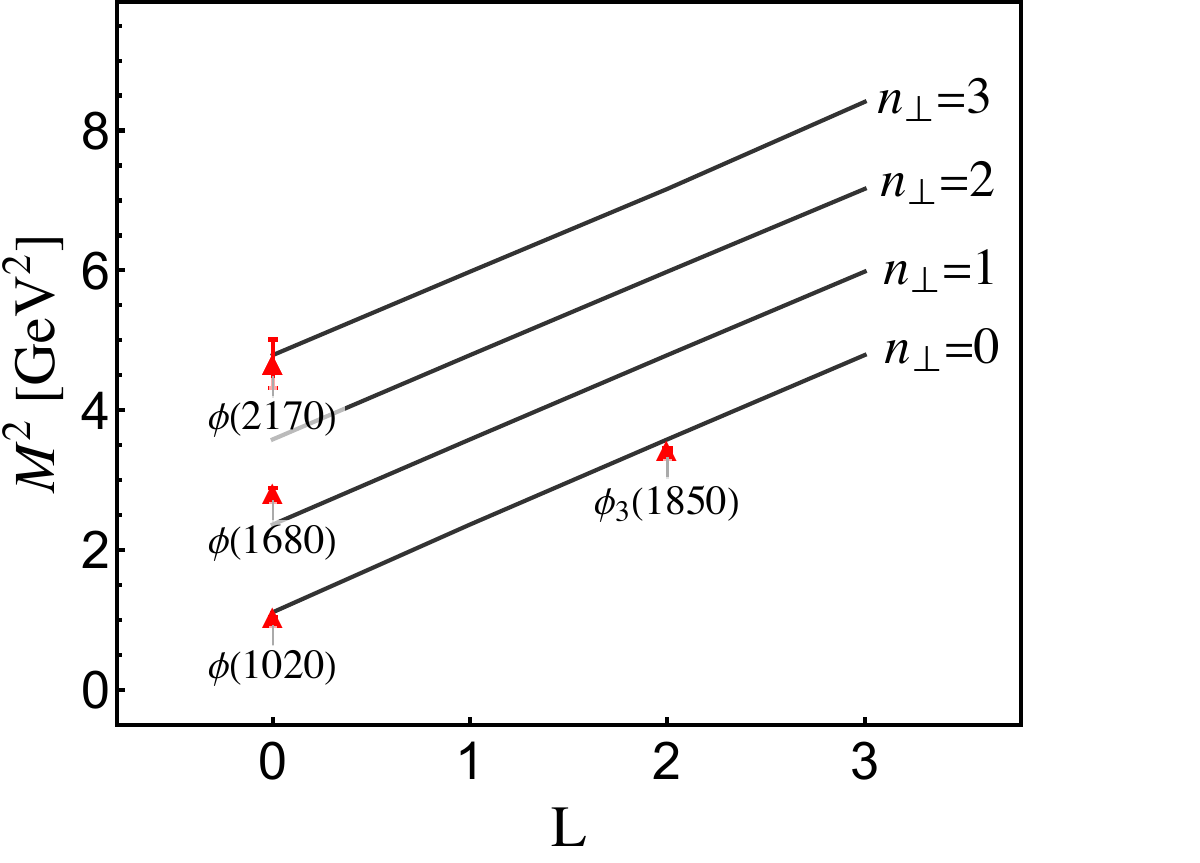}
    \caption{Our predicted Regge trajectories for $\phi$-meson family in comparison with the experimental data~\cite{ParticleDataGroup:2018ovx}.}
    \label{fig:mass-spectra}
\end{figure}
%%%%%%%%%%%%%%

Brodsky and de Téramond proposed a nonperturbative method to calculate the spin-independent component of hadronic LFWFs based on a semiclassical approximation of light-front QCD~\cite{PhysRevLett.94.201601,PhysRevLett.96.201601,PhysRevLett.102.081601,Brodsky:2014yha}. In this approach, the spin-independent wave function is expressed in a factorized form involving the variables $x,~\zeta,~\textrm{and}~\varphi$ as,
\begin{equation}
	\Psi (x, \zeta, \varphi)= \frac{\phi (\zeta)}{\sqrt{2\pi \zeta}} e^{i L \varphi} X(x),
	\label{full-mesonwf}
\end{equation}
where  $\zeta = \sqrt{x(1-x)}r_{\perp}$ connects LFWFs to the AdS$_5$ space. The transverse mode, $\phi(\zeta)$, contains the dynamical properties of the hadronic LFWF. The light-front orbital angular momentum quantum number is denoted by $L$, and the longitudinal wave function is given by
$
X(x) = \sqrt{x(1-x)}\chi(x).
$ 
The dynamics of transverse modes can be generated with the holographic light-front Schr\"{o}dinger equation~\cite{Brodsky:2006uqa,deTeramond:2005su,deTeramond:2008ht,Brodsky:2014yha},
\begin{equation}
	\left(-\frac {\mathrm{d}^2}{\mathrm{d} \zeta^2}+\frac{4L^2-1}{4 \zeta^2}+U_\perp(\zeta)\right)\phi(\zeta)= M_\perp^2 \phi(\zeta),
	\label{SEq}
\end{equation}

where the confining potential, $U_{\perp}$ is given as a two-dimensional (2D) harmonic oscillator potential in holographic variable $\zeta$ as,
\begin{equation}
	U_\perp(\zeta)=\kappa^4 \zeta^2 + 2\kappa^2(J-1),
	\label{U-LFH}
\end{equation}
with, $J=L+S$, the total angular momentum of the meson. Eq.~(\ref{SEq}) can be solved analytically by employing a holographic mapping to AdS$_{5}$ through the light-front variable $\zeta$, within the underlying conformal symmetry framework, as discussed in Ref.~\cite{Brodsky:2013ar}. Here, the emerging mass scale parameter $\kappa$ fixes the confinement strength and provides the meson mass spectra, as well as the corresponding meson eigenstate within the chiral limit as:
\begin{equation}
	M_{\perp}^2(n_\perp , J, L)=4\kappa^2\left(n_\perp + \frac{J+L}{2}\right),
	\label{MTM}
\end{equation}
and 
\begin{equation}
	\phi_{n_{\perp} L}(\zeta )\propto \zeta^{1/2+L}\exp\left(\frac{-\kappa^2\zeta^2}{2}\right)L_{n_\perp}^L(\kappa^2\zeta^2),
	\label{TMWF}
\end{equation}
where $n_{\perp}$ is the transverse principle quantum number, $L_{n_\perp}^{L}$ represents the associated Laguerre polynomials. Within the chiral limit, the lowest-lying hadronic bound state is massless pion with, $n_{\perp}=L=S=0$.
\begin{figure}
		\centering	
    \includegraphics[scale=0.55]{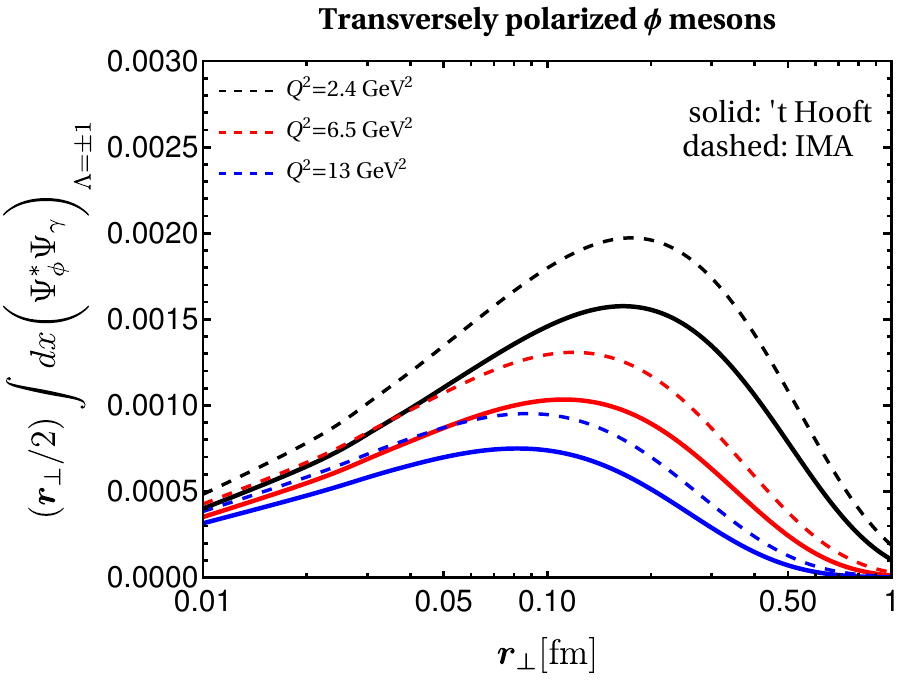}
    \hspace{0.5cm}
	\includegraphics[scale=0.55]{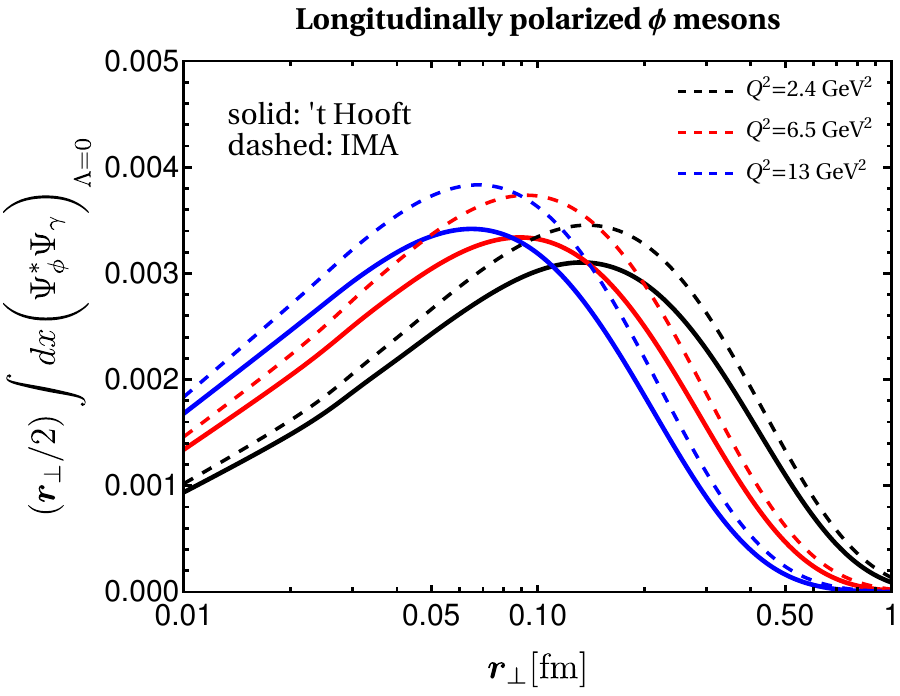}
    \caption{The overlap functions of both the transverse ($\Lambda=\pm 1$) and the longitudinal ($\Lambda=0$) photon and the $\phi$-meson LFWFs, are integrated over $x$ and presented as a function of the dipole transverse size ${\mathbf r}_{\perp}$ (in fm) at different photon virtualities ($Q^2 = 2.4, 6.5$, and $13$ GeV$^2$). These predictions are provided by the light-front holography 't Hooft (solid) and the light-front holography IMA (dashed) approaches, respectively.}
	\label{Fig:LFWFsoverlap}
\end{figure}

Beyond the chiral limit, Brodsky and de Téramond proposed a prescription to describe the longitudinal mode based on invariant mass as~\cite{Brodsky:2008pg}:

\begin{align}
	X(x) &= \sqrt{x(1-x)}\exp\left[-\frac{1}{2\kappa^2}\left(\frac{m_q^2}{x}+\frac{m_{\bar{q}}^2}{1-x} \right)\right].
	\label{bda}
\end{align}
Note that this longitudinal mode is not dynamical. The corresponding meson mass spectra in Eq.~\eqref{MTM} becomes,
\begin{align}
    M^{2}=\Delta M^{2}+4\kappa^2\left(n_\perp + \frac{J+L}{2}\right),
    \label{total mass}
\end{align}
where the mass shift $\Delta M$ is obtained from the first order corrections as follows:
\begin{align}
   \Delta M^2 = \int \frac{\mathrm{d} x}{x(1-x)} X^2(x) \left(\frac{m_q^2}{x}+\frac{m_{\bar{q}}^2}{1-x} \right).
	\label{massshift}	
\end{align} 
Hence for the lowest lying mesonic state, Eq.~\eqref{total mass}, implies that: $\Delta M=M_{\pi}$.
It is important to note that this prescription suffers from two shortcomings: (i) $M_{\pi}^2 \propto 2m_q^2 (\ln(\kappa^2/m_q^2)-\gamma_E)$
with $\gamma_E=0.577216$, which contradicts the Gell-Mann-Oakes-Renner (GMOR) relation, $M_{\pi}^2\propto m_q$~\cite{Ahmady:2022dfv,Gurjar:2024wpq}.  (ii) The longitudinal mode, Eq.~\eqref{bda}, remains same for all the radially excited states. However, despite these issues, this prescription has been successfully employed to describe various properties of both light and heavy mesons
~\cite{Gurjar:2024wpq,Brodsky:2014yha,Brodsky:2008pg,deTeramond:2021yyi,Swarnkar:2015osa,Ahmady:2020mht,Ahmady:2019hag,Ahmady:2019yvo,Ahmady:2018muv,Ahmady:2016ufq}.
\begin{figure}
		\centering	
	\includegraphics[scale=0.68]{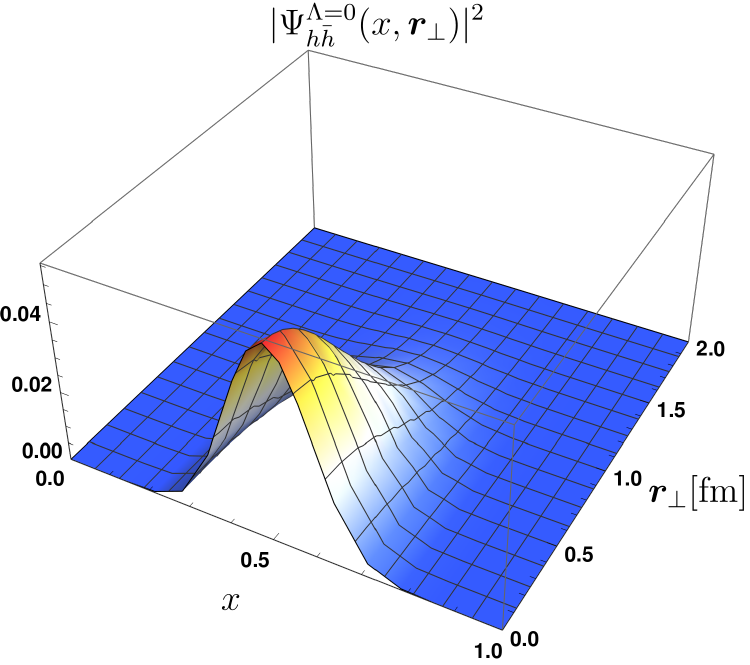}\hspace{0.5cm}
	\includegraphics[scale=0.68]{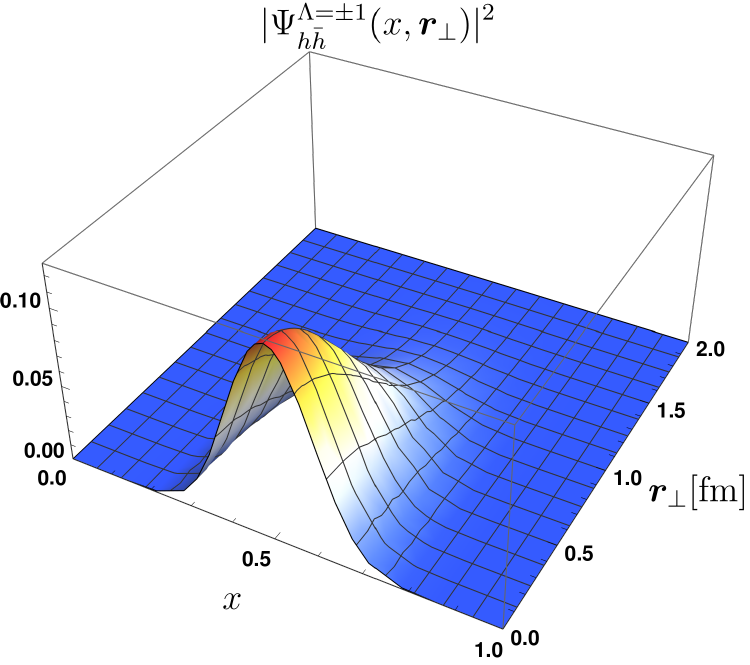}
    \caption{The three-dimensional distribution of longitudinally (left) and transversely (right) polarized LFWFs of the $\phi$-meson as a functions of the longitudinal momentum fraction carried by the quark $x$ and the dipole separation $\mathbf{r}_{\perp}$ (in fm).}
    \label{Fig:3DLFWFs}
\end{figure}

In Refs.~\cite{Ahmady:2021lsh,Ahmady:2021yzh} the longitudinal dynamics has been captured by the 't Hooft equation~\cite{tHooft:1974pnl}  of (1 + 1)-dim at large $N_c$ QCD, which describe the full meson's mass spectra for pion~\cite{Ahmady:2022dfv}, $\rho$ meson~\cite{Gurjar:2024wpq}, as well as heavy-light and heavy-heavy mesons~\cite{Ahmady:2021yzh}. The idea of utilizing the 't Hooft equation to go beyond the invariant mass prescription was first proposed in Ref.~\cite{Chabysheva:2012fe}, which focus to predict meson's decay constants and parton distribution functions. Recent studies~\cite{deTeramond:2021yyi,Li:2021jqb} have employed a phenomenological longitudinal confinement potential, originally proposed in Ref.~\cite{Li:2015zda} within the framework of basis light-front quantization (BLFQ). While both studies primarily focus on the chiral limit and the phenomenon of chiral symmetry breaking, the investigation in~\cite{deTeramond:2021yyi} extends to heavy mesons in their ground state and examines the relationship of their method to the 't Hooft equation. It is worth noting that there is a growing interest in integrating longitudinal dynamics into hLFQCD~\cite{Li:2021jqb,deTeramond:2021yyi,Lyubovitskij:2022rod,Weller:2021wog,Ahmady:2021lsh,Rinaldi:2022dyh}.

The 't Hooft equation can be derived by applying the QCD Lagrangian in (1+1)-dimensions with large $N_{c}$ approximations, as shown in~\cite{tHooft:1974pnl},
\begin{eqnarray}
\left(\frac{m_{q}^{2}}{x}+\frac{m_{\bar{q}}^{2}}{1-x}\right)\chi(x)+\frac{g^{2}}{\pi}\mathcal{P}\int {\mathrm d}y \frac{\chi(x)-\chi(y)}{(x-y)^{2}}=M^{2}_{\parallel}\chi(x),
\label{eq:tHooft}
\end{eqnarray}
where $g=g_{s}\sqrt{N_{c}}$ indicates the 't Hooft longitudinal confinement scale and $\mathcal{P}$ is the Cauchy principal value. The 't Hooft equation possesses a gravity dual on AdS$_{3}$~\cite{Katz:2007br} and has gained significant attention in the literature~\cite{Zhitnitsky:1985um,PhysRevD.57.1366,PhysRevLett.69.1018,PhysRevD.103.074002,PhysRevD.62.094011,PhysRevD.104.036004,Ahmady:2021lsh,Ahmady:2022dfv,PhysRevD.104.074013}. Unlike the holographic light-front Schr\"odinger equation, the 't Hooft equation cannot be solved analytically. However, it can be tackled numerically using the matrix method as described in Ref.~\cite{Chabysheva:2012fe}. After including both the transverse modes from holographic light-front Schr\"odinger equation and longitudinal modes from 't Hooft equation, the total meson mass spectra can be obtained as,
\begin{align}
	M^2(n_\perp ,n_\parallel ,J, L)= 4\kappa^2\left(n_\perp + \frac{J+L}{2}\right)+ M_\parallel^2(n_\parallel , m_q, m_{\bar q}, g),
	\label{totalmass}
\end{align}
where, $n_{\parallel}$ is the longitudinal quantum number. For the ground state ($n_{\perp}=L=S=0$ and $n_{\parallel}=0$), the above equation reproduces the experimentally observed pion mass, whereas it is zero in Eq.~(\ref{MTM}). This indicates that the 't Hooft equation generates the total pion mass. We also find that both the holographic Schr\"odinger and the 't Hooft equations correctly predicts the GMOR relation $M_{\pi}^{2}\sim m_{q}$~\cite{deTeramond:2021yyi,Ahmady:2022dfv}. Although an exact analytical expression for the longitudinal modes cannot be obtained, they can be approximately fitted to the following form: 
\begin{eqnarray}
\chi(x)\approx x^{\beta_{1}}(1-x)^{\beta_{2}},
\label{eq:tHooftanaly}
\end{eqnarray}
where $\beta_i$ are the quark mass dependent parameters which vanishes in the chiral limit. Specifically, for the ground state of the $\phi$-meson, the obtained values are $\beta_{1,2}\approx6.0$. 
After incorporating both the transverse modes from the light-front Schrödinger equation and the longitudinal modes from the 't Hooft equation, the total spin-independent part of the meson LFWFs can be expressed as follows:
\begin{eqnarray}
\Psi (x,\zeta) = \mathcal{N} \sqrt{x (1-x)}
\chi(x)\exp{ \left[ -{ \kappa^2 \zeta^2  \over 2} \right] },
\label{eq:totalhwf}
\end{eqnarray}
where $\chi(x)$ is the longitudinal mode, which is obtained by solving the ’t Hooft equation.  $\mathcal{N}$ is the  normalization constant, which can be fixed by using the normalization condition as,
\begin{eqnarray}
\sum_{h,\bar{h}} \int {\mathrm d}^2 {\mathbf{r}_{\perp}} \, {\mathrm d} x |
\Psi^{V, \Lambda} _{h, {\bar h}}(x, {\mathbf r}_{\perp})|^{2} = 1 ,
\label{eq:normalization}
\end{eqnarray}
where $\Psi_{h,\bar{h}}^{V,\Lambda}$ are the spin-improved wave functions and these are given in Eqs.~(\ref{eq:LFWFlong}) and (\ref{eq:LFWFtrans}). 

\begin{figure}
		\centering	
	\includegraphics[scale=0.55]{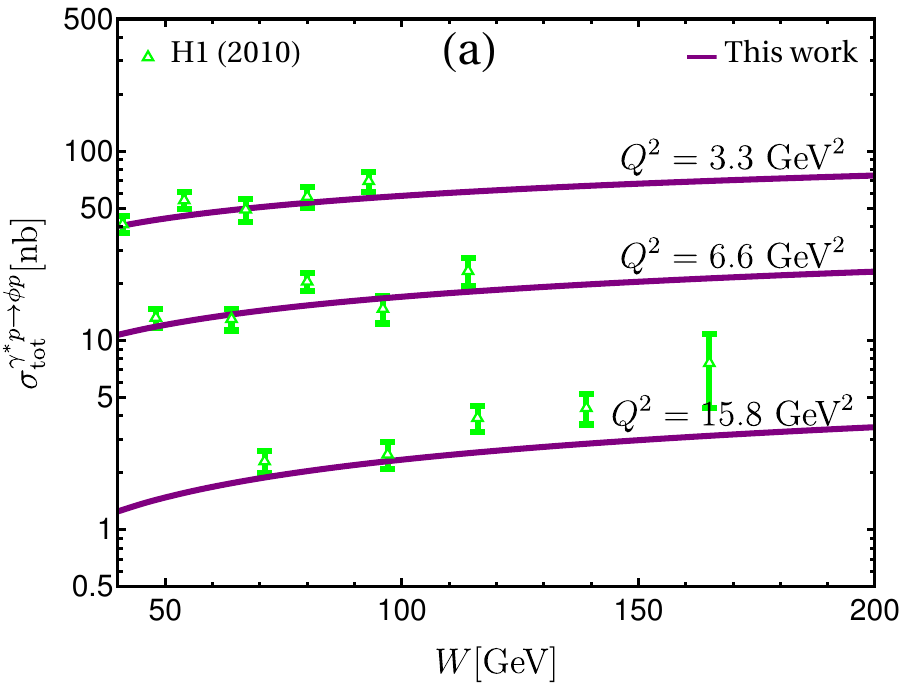}\hspace{0.5cm}
	\includegraphics[scale=0.55]{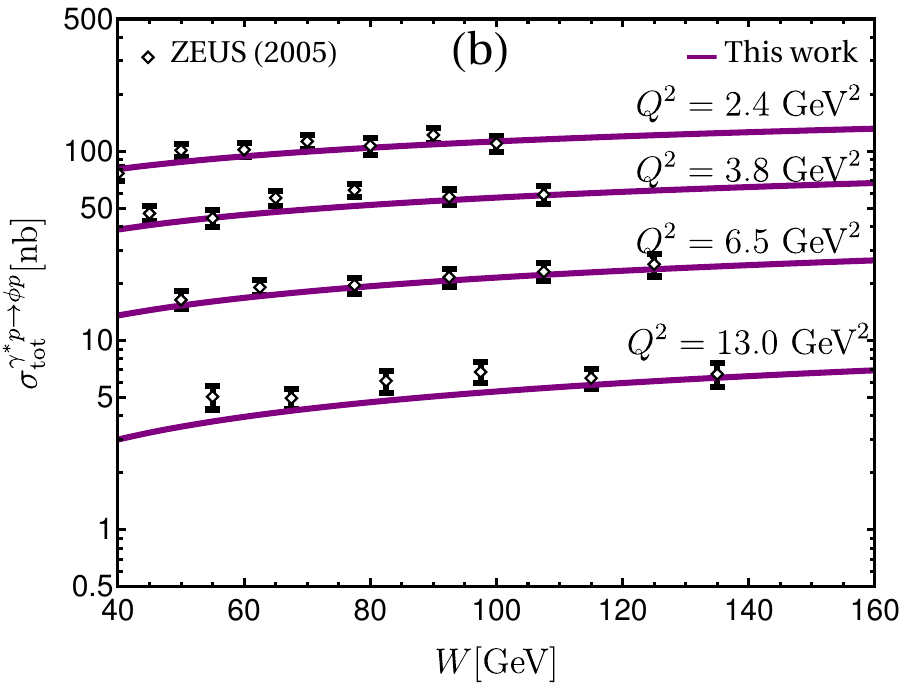}
    \caption{Comparison of our predictions for total diffractive cross-section 
    as a function of photon-proton center of mass energy $W$ (in GeV) for various $Q^{2}$ bins (in GeV$^2$) with experimental data from (a) H1 2010~\cite{H1:2009cml} and (b) ZEUS 2005~\cite{ZEUS:2005bhf}.}
	\label{Fig:totalcrossW}
\end{figure}

%========================
\section{Results and discussion}\label{sec:results}
%========================
%========================
\subsection{Mass spectroscopy}
%=====================
The parity and charge conjugation quantum numbers of meson states are determined by~\cite{Ahmady:2021lsh,Ahmady:2021yzh,Ahmady:2022dfv}:
\begin{align}
    P=(-1)^{L+1}\,, ~~~~~~~~~~~~~~C=(-1)^{L+S+n_{\parallel}}.
\end{align}
Using the universal transverse confinement scale $\kappa = 0.523$ GeV and the strange quark mass $m_{s} = 0.357$ GeV, values adopted in hLFQCD along with the IMA~\cite{Brodsky:2014yha}, we apply the longitudinal confining scale $g = 0.109$ GeV~\cite{Ahmady:2022dfv} to extract the mass spectrum for the $\phi$-meson family and determine their corresponding wave functions.
We present our computed masses for the $\phi$-meson and its excited states in Table~\ref{Spectroscopy}. Our results (last column) show good agreement with experimental data (second column, in parentheses). 
Notably, the condition $n_\parallel \ge n_\perp + L$ observed in Table~\ref{Spectroscopy} holds true across the full hadron spectrum~\cite{Ahmady:2021yzh}. The Regge trajectories for the $\phi$-meson family resulting from our calculations are displayed in Fig.~\ref{fig:mass-spectra}.
%%%%%%%%%%%%%%%%%%%%%%%%%%%%%%%%%%%%%%%
\subsection{$\phi$-meson diffractive cross-section}
%%%%%%%%%%%%%%%%%%%%%%%%%%%%%%%%%%%%%%%
The diffractive cross-section, Eq.~(\ref{eq:amplitude-VMP}), can be expressed as the overlap of the LFWFs of the virtual photon and vector meson. In Fig.~\ref{Fig:LFWFsoverlap}, we show the overlap of the LFWFs after integrating over the longitudinal momentum fraction $x$ for various photon virtualities: $Q^{2} = 2.4,~6.5,$ and $13~\mathrm{GeV}^{2}$. We compare the 't Hooft overlap functions, which incorporate the longitudinal dynamics, with the IMA overlap functions, which do not include longitudinal dynamics in the meson wave functions. Interestingly, despite their theoretical differences, these functions exhibit analogous behavior. 
Specifically, {the peaks of the distributions for the overlap function of transversely polarized photons and $\phi$ mesons} shift towards smaller values of the transverse separation ${\mathbf r}_{\perp}$ and decrease in magnitude as the photon virtuality increases, {whereas for longitudinally polarized photons and $\phi$ mesons, the peaks also shift towards smaller transverse separations, but their magnitude increases with increasing photon virtuality}.
\begin{figure}
		\centering	
\includegraphics[scale=0.55]{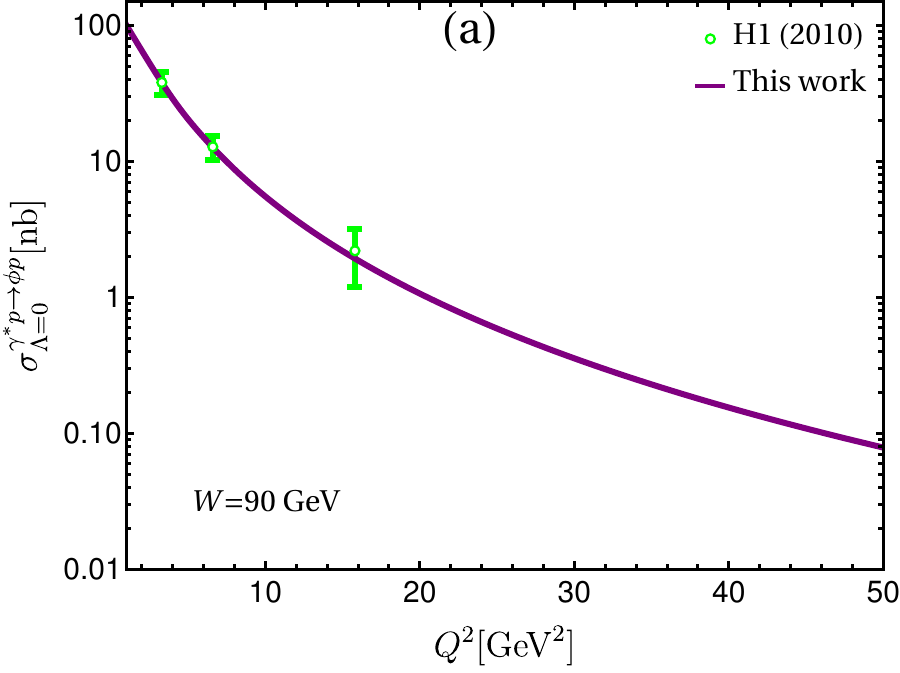}\hspace{0.5cm}
	\includegraphics[scale=0.55]{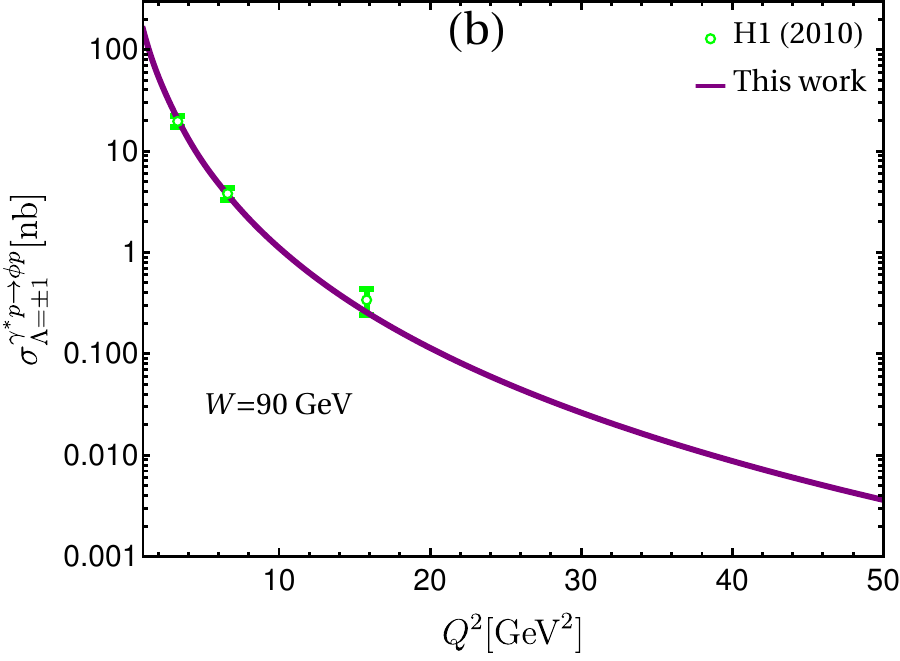}
    \vspace{0.5cm}
 \includegraphics[scale=0.55]{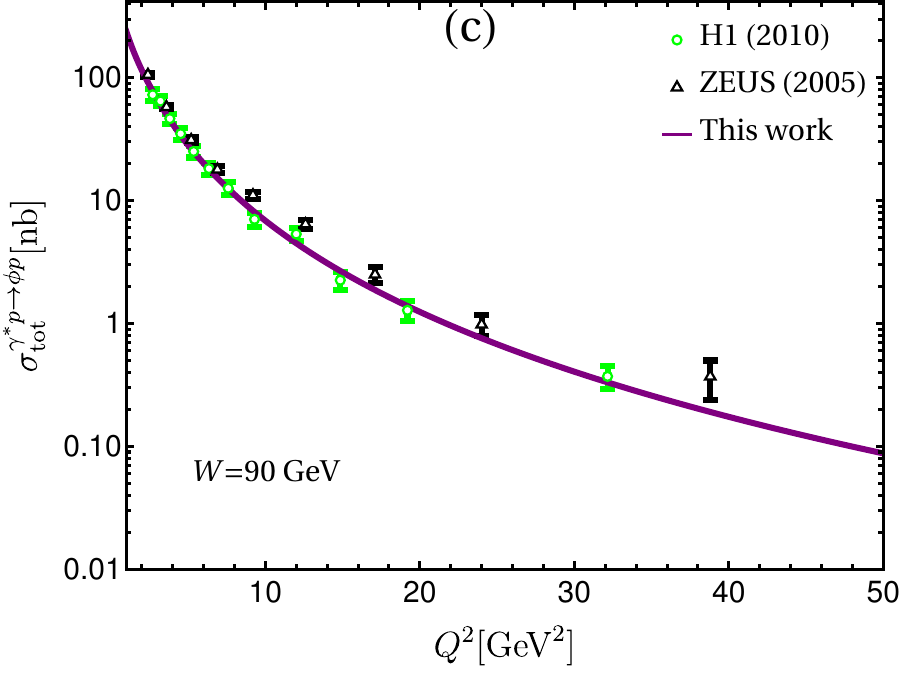}\hspace{0.5cm}
	\includegraphics[scale=0.65]{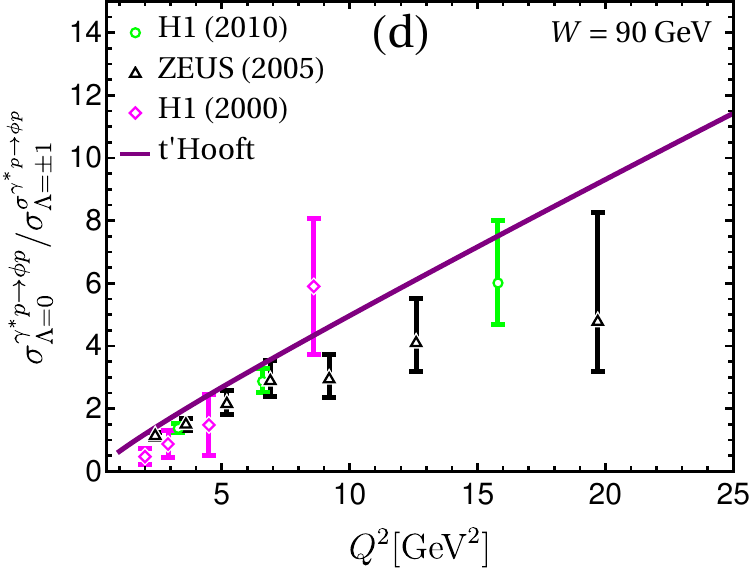}
    \caption{Comparison of our predictions as a function of $Q^{2}$ for (a) longitudinal, (b) transverse, (c) total $\gamma^{\ast}p$ cross-sections with $W = 90$ GeV and (d) longitudinal to transverse cross-section ratio with H1 at $W=90$ GeV~\cite{H1:1999pji,H1:2009cml} and ZEUS at $W=75$ GeV~\cite{ZEUS:2007iet} data.}
	\label{Fig:totalcross}
\end{figure}

In Fig.~\ref{Fig:3DLFWFs}, we present the three-dimensional probabilistic distributions of the $\phi$-meson LFWFs, $|\Psi_{h,\bar{h}}^\Lambda (x,{\mathbf r}_{\perp})|^2$, as a function of the longitudinal momentum fraction $x$ and dipole transverse separation ${\mathbf r}_{\perp}$, for both longitudinally and transversely polarized $\phi$-mesons. Our holographic LFWFs, incorporating longitudinal modes generated by the 't Hooft equation, peak at $x = 0.5$ and ${\mathbf r}_{\perp} = 0$, and smoothly approach zero as $x \to 0, 1$ and ${\mathbf r}_{\perp}$ increases. Additionally, the transverse wave function exhibits a {broader distribution} compared to the longitudinal wave function. The $\phi$ wave functions qualitatively resemble the $\rho$ wave functions~\cite{Gurjar:2024wpq}, though with a slightly sharper peak.

In Fig.~\ref{Fig:totalcrossW}, we show the variation of total cross section for the $\gamma^{\ast}p \rightarrow \phi p$ process as a function of photon-proton
center of mass energy $W$ for different $Q^{2}$ bins. We compare our predictions with data from H1~\cite{H1:2009cml} (left panel) and ZEUS~\cite{ZEUS:2005bhf} (right panel), and observe that our results are in good agreement with the experimental data within the allowed uncertainty range.
Meanwhile, in Figs.~\ref{Fig:totalcross}(a) and \ref{Fig:totalcross}(b), we present the longitudinal ($\Lambda=0$) and transverse ($\Lambda=\pm1$) components of the $\gamma^{\ast}p \rightarrow \phi p$ cross sections as a function of the photon virtuality $Q^{2}$ at a fixed value of $W=90$ GeV. In Fig.~\ref{Fig:totalcross}(c), the total cross section for the $\gamma^{\ast}p \rightarrow \phi p$ process is shown as a function of $Q^{2}$. Finally, Fig.~\ref{Fig:totalcross}(d) depicts the ratio of the longitudinal to transverse components of the cross section. We find a good consistency between our model results and the experimental data across all cases.

\begin{figure}
		\centering	
	\includegraphics[scale=0.55]{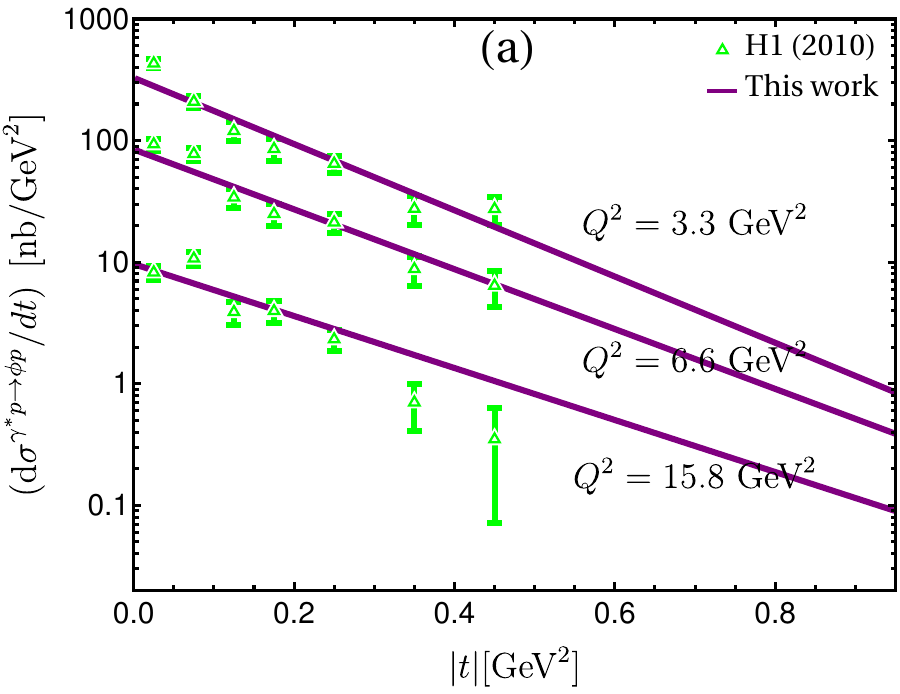}\hspace{0.5cm}
	\includegraphics[scale=0.55]{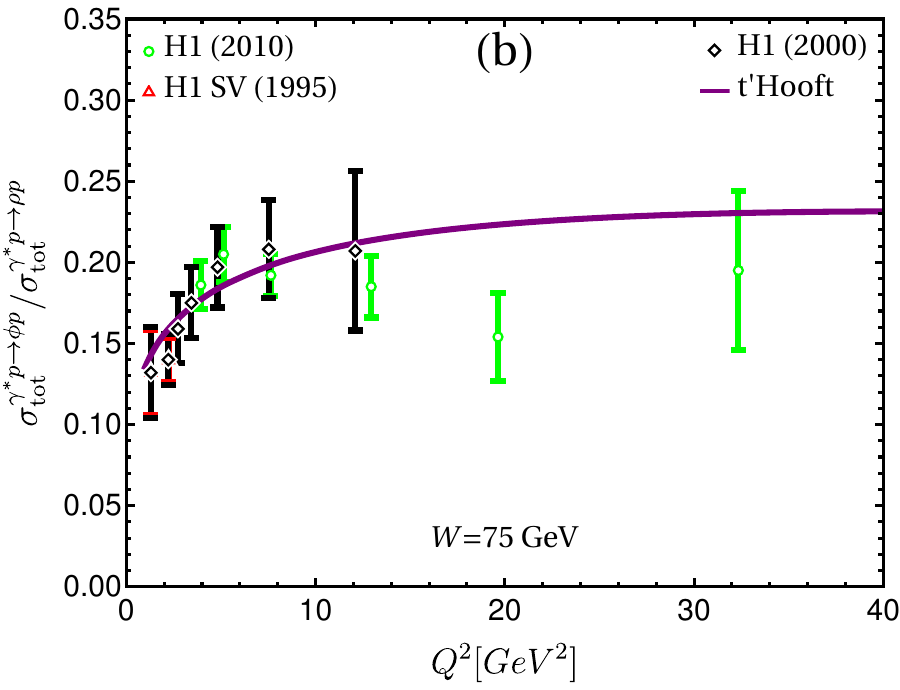}
    \caption{Comparison of our predictions for differential cross-section as a function of $|t|$ for various $Q^{2}$ bins (left) and the ratio for $\phi$ to $\rho$ production total cross-section as a function of $Q^{2}$ with H1 data at $W=75$ GeV~\cite{H1:1999pji,H1:2009cml} (right). The $\rho$-meson production cross section results are taken from our previous work~\cite{Gurjar:2024wpq}.}
	\label{Fig:diffcross}
\end{figure}

In Fig.~\ref{Fig:diffcross}(a), we present the differential cross-section ${\mathrm d}\sigma^{\gamma^{\ast}p\rightarrow \phi p}/{\mathrm{d}t}$ as a function of $|t|$ at specific values of $Q^{2}$, compared with the H1 2010 data~\cite{H1:2009cml}. In Fig.~\ref{Fig:diffcross}(b), we display the ratio of the total cross-section for $\phi$ to $\rho$ as a function of $Q^{2}$, where the $\rho$-meson cross section results are taken from our previous work~\cite{Gurjar:2024wpq}. Our predictions in both cases show a good agreement with the H1 HERA data. Note that if the $\rho$ and $\phi$ mesons have identical masses and LFWFs, the ratio of their cross-sections is simply given by the squared ratio of their effective charges from the quark-antiquark dipole coupling with the photon, i.e., $e_{s}^{2}/e_{u,d}^{2} = 0.22$, and our results are approaching it for large values of $Q^{2}$.

%%%%%%%%%%%%
\subsection{Decay constant and Distribution Amplitudes}
%%%%%%%%%%%%
The vector and tensor coupling constants, $f_V$ and $f_V^T$, are defined as the local matrix elements for the transition from vacuum to hadron~\cite{Ball:1998sk,PhysRevD.91.014018},
\begin{eqnarray}
	\langle 0|\bar q(0)  \gamma^\mu q(0)|V
	(P,\Lambda)\rangle =f_V M_{V} \epsilon_\Lambda^{\mu}\,,
	\label{eq:fv}
\end{eqnarray}
and
\begin{eqnarray}
	\langle 0|\bar q(0) [\gamma^\mu,\gamma^\nu] q(0)|V (P,\Lambda)\rangle =2 f_V^{T}  (\epsilon^{\mu}_{\Lambda} P^{\nu} - \epsilon^{\nu}_{\Lambda} P^{\mu}).
	\label{eq:fvT}
\end{eqnarray}
Here, $q(\bar{q})$ represent the quark (anti-quark) field operators at the same space-time points. The momentum and polarization vectors are denoted by $P^{\mu}$ and $\epsilon_{\Lambda}^{\mu}$, respectively. In terms of LFWFs, the decay constants can be written as~\cite{PhysRevD.87.054013},
\begin{eqnarray}
f_{V} &=&  {\sqrt \frac{N_c}{\pi} }  \int_0^1 {\mathrm d} x  \left[ 1 + { m_{q}^{2}-\nabla_{\mathbf{r}_{\perp}}^{2} \over x (1-x) M^{2}_{V} } \right] \left. \Psi(x,\mathbf{r}_{\perp}) \right|_{\mathbf{r}_{\perp}=0}\,,
\label{eq:fvL}	
\end{eqnarray}
and
\begin{eqnarray}
	f_{V}^{T}(\mu) =\sqrt{\frac{N_c}{2\pi}} m_q \int_0^1 {\mathrm d} x \; \int {\mathrm d} \mathbf{r}_{\perp} \; \mu J_1(\mu \mathbf{r}_{\perp})  \frac{\Psi(x,\mathbf{r}_{\perp})}{x(1-x)}
	\label{eq:fvT1}\,,
\end{eqnarray}
where $N_{c}$ is the number of colors, $\mu$ denotes the ultraviolet cut-off scale, and $J_{1}$ is the Bessel function of the first kind of order one.
We observe that the tensor coupling is scale-independent for $\mu^{2} \geq 1$, but it depends on the quark mass $m_q$, as shown in Eq.~(\ref{eq:fvT1}). In the chiral limit, where $m_q \to 0$, the tensor coupling vanishes, while the vector coupling retains a nonzero value. The vector coupling can be used to calculate the electronic decay width~\cite{ParticleDataGroup:2018ovx},
\begin{eqnarray}
	 \Gamma_{V \rightarrow e^+ e^-}={ 4 \pi  \alpha_{\rm em}^2  C_V^2 \over 3 M_V }f_V^2\,.
	\label{eq:decaywidth}
\end{eqnarray}
Here, $\alpha_{\rm em}$ represents the fine structure constant, and $C_{V} = 1/3$ for the $\phi$-meson.
In Table~\ref{tab:fv-phi}, we compare our predictions for the vector and tensor decay constants of the $\phi$-meson, as well as their ratio ($f_{\phi}^{T}/f_{\phi}$), with those obtained from lattice QCD~\cite{Becirevic:2003pn,Braun:2003jg}, other theoretical approaches~\cite{Ahmady:2019hag,Ball:1998kk,Gao:2014bca}, and available experimental data~\cite{ParticleDataGroup:2018ovx}. We observe that our predictions for the vector and tensor decay constants are lower than those of other models, while their ratio is in good agreement. We obtain the electronic decay width as $\Gamma_{\phi \rightarrow e^+ e^-}=0.55$ keV in the LF holography with longitudinal dynamics generated by 't Hooft equation, while the value in LF holography with  IMA is $0.89$ keV~\cite{PhysRevD.94.074018}. The PDG reports it as $\Gamma_{\phi \rightarrow e^+ e^-}=1.251 \pm 0.021$ keV~\cite{PhysRevD.98.030001}. 

The longitudinal and transverse components of the twist-2 DAs are expressed in terms of the vector and tensor decay constants and the LFWFs as~\cite{PhysRevD.88.014042,Gurjar:2024wpq},
\begin{eqnarray}
	\phi_V^{\|}(x, \mu)=  \frac{N_c}{\pi f_V M_V} \int \mathrm{d} \mathbf{r}_{\perp} \mu J_1(\mu \mathbf{r}_{\perp})\left[M_V^2 x(1-x)\right.
	 \left.+m_q^2-\nabla_{\mathbf{r}_{\perp}}^2\right] \frac{\Psi(x,\mathbf{r}_{\perp})}{x(1-x)}\,,
\end{eqnarray}
and
\begin{eqnarray}
	\phi_V^{\perp}(x, \mu)=\frac{N_c m_q}{\pi f_V^{\perp}} \int \mathrm{d} \mathbf{r}_{\perp} \mu J_1(\mu \mathbf{r}_{\perp}) \frac{\Psi(x,\mathbf{r}_{\perp})}{x(1-x)},
\end{eqnarray} 
respectively. They are normalized as~\cite{Choi:2007yu}
\begin{eqnarray}
	\int_{0}^{1}{\mathrm d}x \phi_{V}^{\|}(x,\mu)=1,~~~\textrm{and}~~~\int_{0}^{1}{\mathrm d}x \phi_{V}^{\perp}(x,\mu)=1\,.
\end{eqnarray}

\begin{figure}
		\centering	
	\includegraphics[scale=0.55]{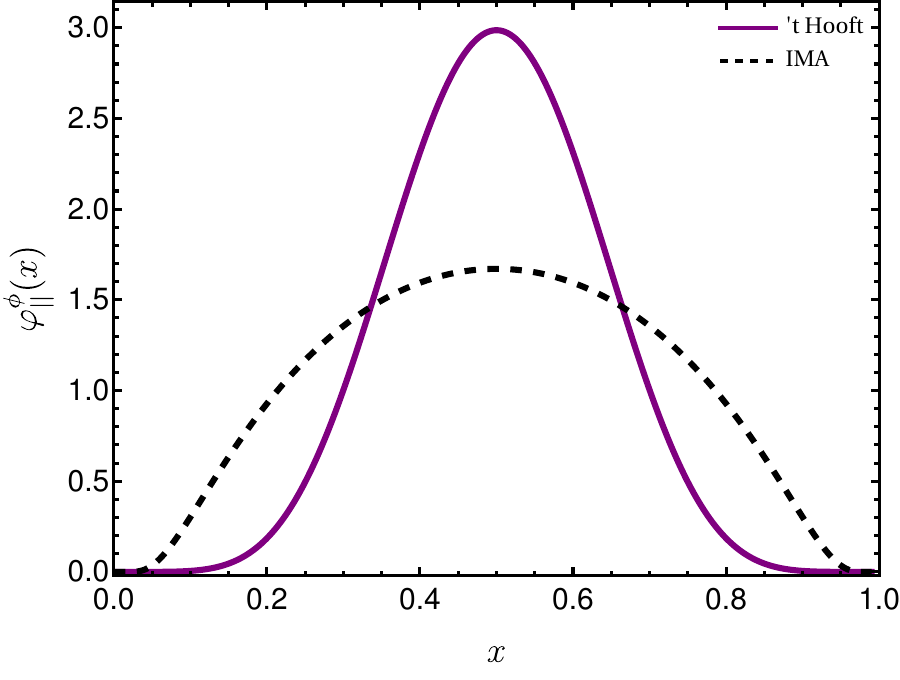}\hspace{0.5cm}
	\includegraphics[scale=0.55]{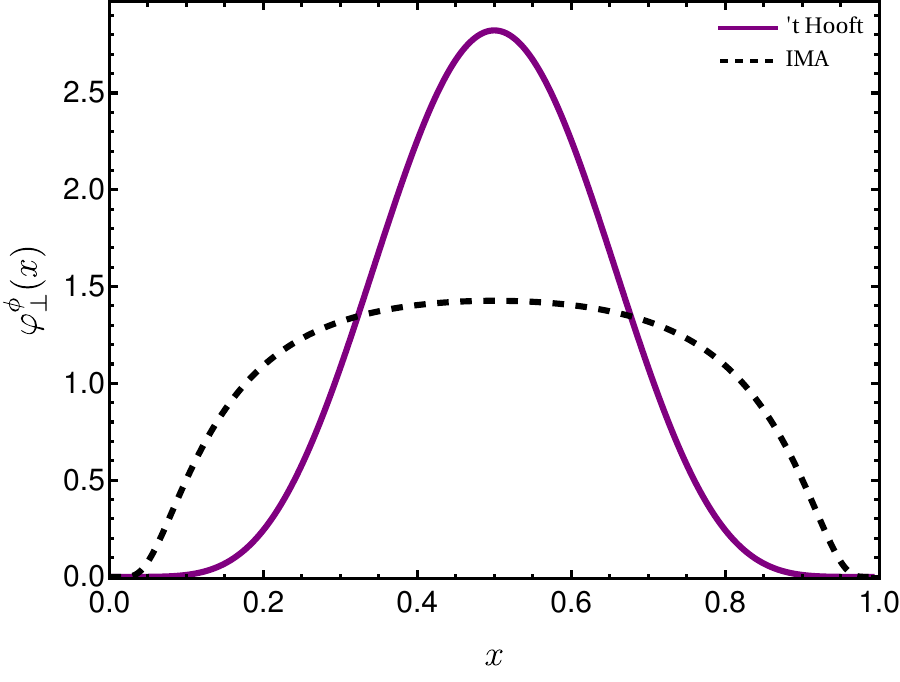}%\vspace{0.5cm}
    \caption{Our results for the PDAs for longitudinally (left) and transversely (right) polarized $\phi$-mesons for 't Hooft (in solid purple) and IMA (black-dashed).}
	\label{Fig:DAs}
\end{figure}

In Fig.~\ref{Fig:DAs}, we present the longitudinal and transverse components of the normalized DAs for the 't Hooft model and compare them with the IMA results for $m_{s}=0.357$ GeV~\cite{Brodsky:2014yha}. 
We observe that both $\phi_{\phi}^{\|}(x)$ and $\phi_{\phi}^{\perp}(x)$ in hLFQCD, incorporating the 't Hooft longitudinal mode, exhibit narrower distributions compared to those obtained using hLFQCD with the IMA framework.

%%%%%
 \begin{table}
 	\caption{Our predictions for the longitudinal and transverse decay constants and their ratio for the $\phi$ meson.} 
 	\label{tab:fv-phi}
 	\centering
 	\begin{tabular}{c|c|c|c|c}
 		\hline\hline
 		Reference & Approach & $f_{\phi}$ [MeV] & $f_{\phi}^{\perp}$ [MeV] & $f_{\phi}^{\perp}/f_{\phi}$\\ \hline
        This work & LF holography 't Hooft & $154$  & $123$ & $0.80$ \\
        \hline
 		Ref.~\cite{Ahmady:2019hag} & LF holography IMA & $190\pm 20$  & $150^{+10}_{-20}$ & $0.79\pm 0.13$\\ \hline
 		PDG \cite{ParticleDataGroup:2018ovx}  & Exp. data  & $225\pm 2$  & $$ & $$\\ \hline
 		Ref. \cite{Ball:1998kk} & Sum Rules & $254 \pm 3$ & $204 \pm 14$ &   \\ \hline
 		Ref. \cite{Becirevic:2003pn} & Lattice (continuum) &  &  & $0.76 \pm 0.01$ \\ \hline
 		Ref. \cite{Braun:2003jg} & Lattice (finite) &  &  & $0.780 \pm 0.008$ \\ \hline
 		Ref. \cite{Gao:2014bca} & Dyson-Schwinger & $190$ & $150$ & $0.79$ \\ \hline
 	\end{tabular}
 \end{table}

%%%%%%%%%
\subsection{Electromagnetic Form Factors}
%%%%%%%%%
The EMFFs of vector mesons can be obtained as the hadronic matrix elements of the electromagnetic current $J^{\mu}$ between the initial and final eigenstates of the vector mesons~\cite{PhysRevC.21.1426},
\begin{align}%\nonumber
    \langle V(P^\prime,\Lambda^\prime)\,|J^{\mu}|\,V(P,\Lambda)\rangle=&-\epsilon^{\ast}_{\Lambda^{\prime}}\cdot\epsilon_{\Lambda}(P+P^{\prime})^{\mu}F_{1}(Q^{2})  +\left(\epsilon^{\mu}_{\Lambda}q\cdot\epsilon^{\ast}_{\Lambda^{\prime}}-\epsilon^{\ast\mu}_{\Lambda^{\prime}}q\cdot\epsilon_{\Lambda}\right)F_{2}(Q^{2})\nonumber\\
    &+\frac{(\epsilon^{\ast}_{\Lambda^{\prime}}\cdot q)(\epsilon_{\Lambda}\cdot q)}{2M_{V}^{2}}(P+P^{\prime})^{\mu}F_{3}(Q^{2})\,.
\end{align}
Here, $q = P^{\prime} - P$ is the momentum transfer, $\epsilon_{\Lambda}$ and $\epsilon_{\Lambda^{\prime}}$ are the polarization vectors for the initial and final state vector mesons, and $F_{1}$, $F_{2}$, and $F_{3}$ are the Lorentz-invariant form factors associated with the physical vector meson mass $M_{V}$. We calculate the hadronic matrix elements in the Breit frame, where the momentum transfer occurs only in one transverse direction, i.e., ($q^{+}=0$, $q_{\perp 1}=Q$, $q_{\perp 2}=0$), with $P_{\perp} = -P_{\perp}^{\prime}$, as defined in Refs.~\cite{Cardarelli:1994yq,PhysRevD.46.2141}:
\begin{eqnarray}%\nonumber
    q^{\mu}=(0,0,Q,0),~~
    P^{\mu}=(M_{V}\sqrt{1+\eta},\,M_{V}\sqrt{1+\eta},\,-Q/2,0)~~
    \text{and}~~
    P^{\prime\mu}=&(M_{V}\sqrt{1+\eta},\,M_{V}\sqrt{1+\eta},\,Q/2,0),
\end{eqnarray}
where $\eta=Q^{2}/4M_{V}^{2}$ is the kinematic factor. We follow the notation $p^{\mu}=(p^{+},p^{-},p^{1},p^{2})$ with $p^{\pm}=p^{0}\pm p^{3}$. 

On the light front, we calculate the form factors using the plus component of the electromagnetic current of quark, $J^+_q(0)$, which can be expressed in terms of its matrix elements as~\cite{PhysRevC.102.055207,Li:2021cwv},
\begin{align}
 I_{\Lambda^\prime, \Lambda}^{+} (Q^2) &\triangleq \langle V(P^\prime,\Lambda^\prime)\left| \frac{J^+_q(0)}{2P^+}\right| V(P,\Lambda) \rangle 
 \nonumber \\
 &= \sum_{h, \bar{h}} \int_0^1\int_0^{\infty} \frac{{\mathrm d}x{\mathrm d}^2\textbf{k}_\perp}{16\pi^{3}}
\Psi^{\Lambda^\prime*}_{h\bar{h}}(x,\mathbf{k}_\perp+(1-x)\textbf{q}_\perp)  \Psi^{\Lambda}_{h\bar{h}}(x,\mathbf{k}_\perp) ,\label{eq:FFs}
\end{align}
By accounting all possible combinations of incoming and outgoing vector meson helicities, $\Lambda, \Lambda^{\prime} = 0, \pm 1$, one can obtain nine matrix elements of the electromagnetic current, $I^{+}_{\Lambda^{\prime},\Lambda}$. These nine elements can be reduced to four: $I_{1,1}^{+},\,I_{1,-1}^{+},\,I_{1,0}^{+}$, and $I_{0,0}^{+}$ by imposing light-front parity and time-reversal invariance~\cite{Cardarelli:1994yq}. In practical computations, instead of directly evaluating the Lorentz-invariant form factors $F_{i}(Q^2)$, it is common to use the physical charge ($G_C$), magnetic ($G_M$), and quadrupole ($G_Q$) form factors, expressed as~\cite{PhysRevD.70.053015,PhysRevD.46.2141},
\begin{align}
    G_{C}=F_{1}+\frac{2}{3}\eta G_{Q}\,, \quad\quad
    G_{M}=-F_{2}\,, \quad \quad
    G_{Q}=F_{1}+F_{2}+(1+\eta)F_{3}.
\end{align}
Note that various prescriptions exist for calculating these types of form factors, such as those proposed by Grach and Kondratyuk (GK)~\cite{Grach:1983hd}, Brodsky and Hiller~(BH)~\cite{PhysRevD.46.2141}, Chung-Coester-Keister-Polyzou (CCKP)~\cite{Chung:1988my}, and Frankfurt-Frederico-Strikman (FFS)~\cite{Frankfurt:1993ut}. In our work, we compute these physical form factors using the BH prescription, which accounts for zero-mode contributions. According to the BH prescription, the form factors are defined in terms of hadronic matrix elements, $I_{\Lambda^{\prime},\Lambda}^{+}$ as,
\begin{equation}
  \begin{aligned}\label{eq:ff_vector}
  G_{\mathrm{C}}^{\rm BH}(Q^2) &= \frac{1}{2P^{+}(1+2\eta)} \left[\frac{3-2\eta}{3}I_{0, 0}^{+} +\frac{16}{3}\eta \frac{I_{1,0}^{+}}{\sqrt{2\eta}}+\frac{2}{3}(2\eta-1)I_{1,-1}^{+}\right ],  \\
  G_{\mathrm{M}}^{\rm BH}(Q^2) &= \frac{2}{2P^{+}(1+2\eta)}\left[I_{0,0}^{+}+\frac{(2\eta-1)}{\sqrt{2\eta}}I_{1,0}^{+}-I_{1,-1}^{+}\right], \\
  G_{\mathrm{Q}}^{\rm BH}(Q^2) &= -\frac{1}{2P^{+}(1+2\eta)}\left[I_{0,0}^{+}-2\frac{I_{1,0}^{+}}{\sqrt{2\eta}}+\frac{1+\eta}{\eta}I_{1,-1}^{+} \right].  
  \end{aligned}
\end{equation}
At zero momentum transfer, these form factors {define the electric charge $e$,} magnetic moment $\mu$ and quadrupole moment $\mathcal{Q}$ as follows~\cite{Hernandez-Pinto:2024kwg,Xu:2019ilh}: 
\begin{align}\label{EQ:STATIC}
    eG_{\mathrm{C}}(Q^{2}=0)=e\,,\quad\quad
    G_{\mathrm{M}}(Q^{2}=0)=\mu\,,\quad\quad
    G_{\mathrm{Q}}(Q^{2}=0)=&\mathcal{Q}.
\end{align}
Meanwhile the charge root-mean-squared (rms) radius of the vector meson can be determined by~\cite{Chung:1988my}
\begin{equation}
  \begin{aligned}\label{eq:ff_obs}
  \langle r^2_{\phi} \rangle &=  -\frac{6}{G_{\mathrm{C}}(0)}  \lim_{Q^2 \to 0}\frac{\partial G_{\mathrm{C}}(Q^2)}{\partial Q^2}.
  \end{aligned}
\end{equation}

\begin{figure}
		\centering	
	\includegraphics[scale=0.42]{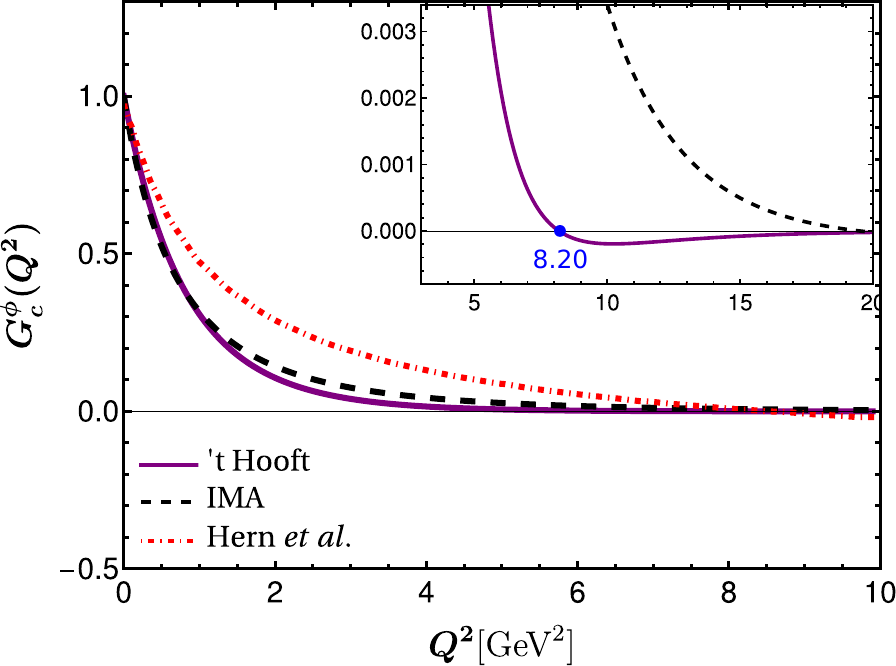}
	\includegraphics[scale=0.405]{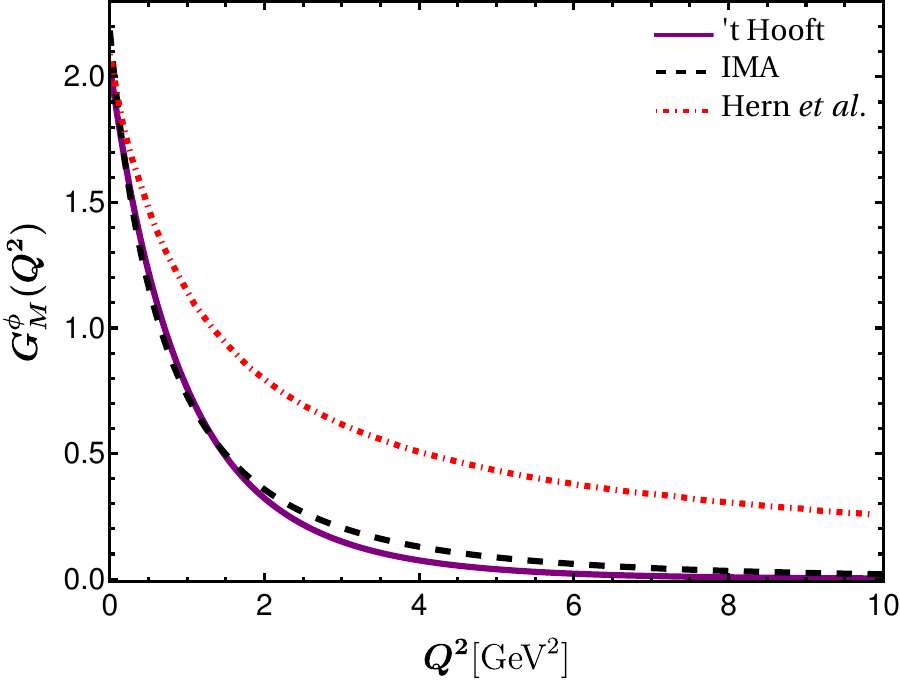}
    \includegraphics[scale=0.41]{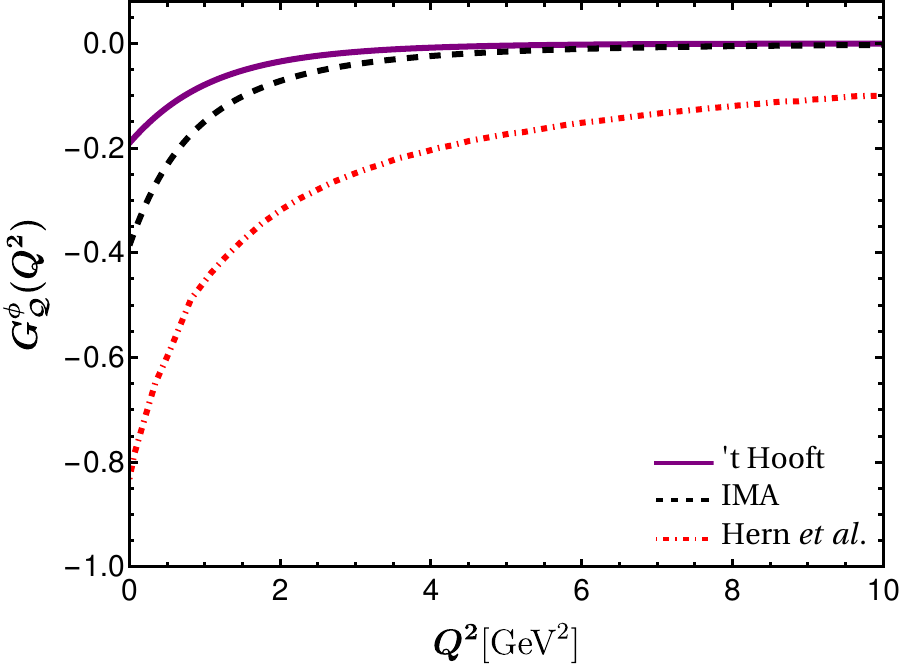}	
    \caption{The left, middle, and right panels show the variation of the electric, magnetic, and quadrupole form factors of $\phi$ mesons as functions of momentum transfer squared, $Q^{2}$. The solid purple line represents the results from 't Hooft, the dashed black line corresponds to the IMA predictions, and the red dot-dashed line represents the Schwinger-Dyson equation results~\cite{Hernandez-Pinto:2024kwg}.}
	\label{Fig:EMFFs}
\end{figure}

In Fig.~\ref{Fig:EMFFs}, we present the variations of the charge, magnetic, and quadrupole electromagnetic form factors as functions of the square of the momentum transfer, $Q^2$. We compare the results obtained from hLFQCD, incorporating the 't Hooft longitudinal mode, with those from hLFQCD within the IMA framework and the Schwinger-Dyson equations approach~\cite{Hernandez-Pinto:2024kwg}. From this comparison, we find that the predictions from the 't Hooft and IMA approaches are largely consistent with each other; however, they fall more rapidly than those obtained using the Schwinger-Dyson equation approach. We observe a zero crossing of the charge form factor, $G_c(Q^2)$, at $Q^2 = 8.2\, \text{GeV}^2$ for hLFQCD with the 't Hooft longitudinal mode, whereas in the IMA approach, it occurs at a higher $Q^2 \sim 20 \, \text{GeV}^2$. Notably, the zero crossing of $G_c(Q^2)$ has also been predicted by the Schwinger-Dyson equation approach, where it appears at $Q^2 = 8.5 \, \text{GeV}^2$~\cite{Hernandez-Pinto:2024kwg}. 

Using Eqs.~\eqref{EQ:STATIC} and \eqref{eq:ff_obs}, we calculate the $\phi$-meson rms charge radius $\sqrt{\langle r_{\phi}^{2}\rangle}$, magnetic moment $\mu_{\phi}$, and quadrupole moment $\mathcal{Q}_{\phi}$. These results are presented alongside other model predictions in Table~\ref{tab:ff_obs1}. Our findings indicate that the charge radius and magnetic moment of the $\phi$-meson are consistent with existing theoretical predictions from the Schwinger-Dyson equation approach~\cite{Hernandez-Pinto:2024kwg} and the symmetry-preserving approach~\cite{Xu:2019ilh}. Additionally, we observe that the quadrupole moment qualitatively aligns with those theoretical predictions as summarized in Table~\ref{tab:ff_obs1}.

\begin{table*}
	\caption{Comparison of the $\phi$-meson charge radii $\sqrt{\langle r^2_{\phi} \rangle}$ (in $\text{fm}$), magnetic moments $\mu_{\phi}$ 
    and quadrupole moment $\mathcal{Q}_{\phi}$ 
    with various theoretical approaches.} 
	\begin{ruledtabular}
		\begin{tabular}{c@{\hskip 0.05in} c@{\hskip 0.05in} c@{\hskip 0.05in}  c@{\hskip 0.05in} c@{\hskip 0.05in}}
			\text{Static properties} & \multicolumn{2}{c}{{\text{This work}}} & Ref.~\cite{Hernandez-Pinto:2024kwg}
			& Ref.~\cite{Xu:2019ilh}\\ 
            \cline{2-3}
                  & 't Hooft     & IMA    &      \\ \hline
$\sqrt{\langle r^2_{\phi}\rangle}$ & $0.54$
            & $0.60$     &   $0.47$    &  $0.52$ \\ 
$\mu_\phi$ & $2.04$
            & $2.18$    &$2.09$  & $2.08$      \\ 
$\mathcal{Q}_\phi$  & $-0.19$
            &$-0.38$   &   $-0.83$  & $-0.32 $
		\end{tabular}
	\end{ruledtabular}
	\label{tab:ff_obs1}
\end{table*}

%%%%%%%%%%%%%%%%%%%%%%%%%%%%%%%%%%%%%%%
%%%%%%%%%%%%%%%%%%%%%%%%%%%%%%%%%%%%%%%
{The matrix elements of the electromagnetic current, $J^{+}$, for spin-1 particles satisfy the angular condition equation as a constraint in the light-front spin basis, expressed as~\cite{Choi:2004ww}:
\begin{align}\label{eq:Ang_con}
    \Delta(Q^{2})=(1+2\eta)I^{+}_{+1,+1}+I^{+}_{+1,-1}-\sqrt{8\eta}I^{+}_{+1,0}-I^{+}_{0,0}=0.
\end{align}
\begin{figure}
		\centering	
	\includegraphics[scale=0.5]{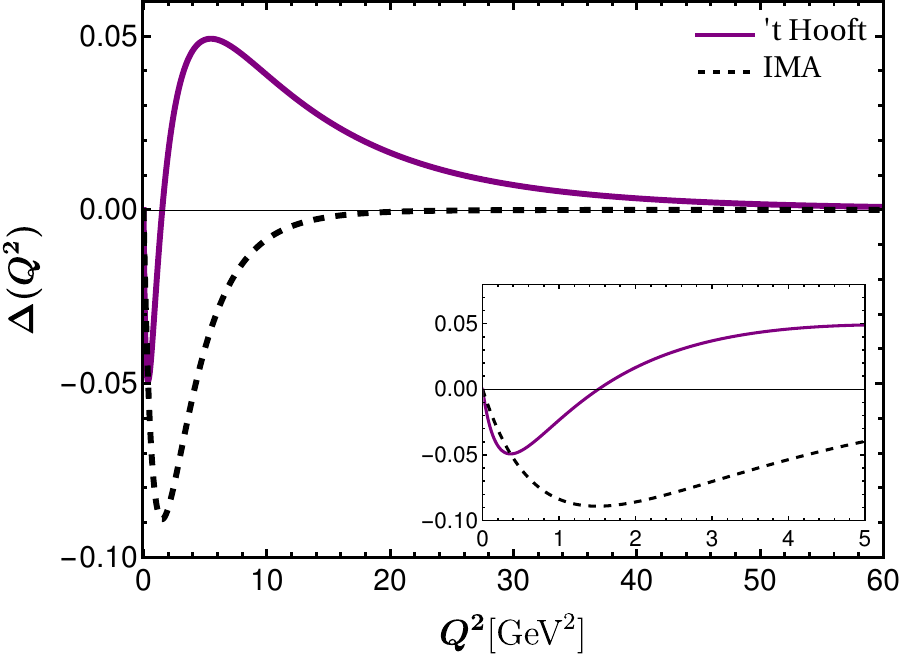}
    \caption{Our results for angular condition given by Eq.~\eqref{eq:Ang_con} for 't Hooft and IMA approaches. The inset depicts the variation of angular condition in the small $Q^{2}$ region where the angular condition is satisfied.} 
	\label{Fig:EMFFs2}
\end{figure}
In Fig.~\ref{Fig:EMFFs2}, we present the variation of the angular condition, Eq.~\eqref{eq:Ang_con}, with the momentum transfer $Q^{2}$. We find that the angular condition constraint $\Delta(0) \rightarrow 0$ is satisfied in our model, suggesting that the contribution from $I^{+}_{0,0}$ is exactly canceled by $I^{+}_{+1,+1}$ at zero momentum transfer, i.e., $I^{+}_{0,0}(0) = I^{+}_{+1,+1}(0)$. However, this equality does not hold across the entire range of $Q^{2}$, although the angular condition eventually approaches zero as $Q^{2}$ becomes very large.
}
%%%%%%%%%%%%%%%%%%%%%%%%%%%%
\section{Conclusion}\label{sec:conclusion}
%%%%%%%%%%%%%%%%%%%%%%%%%%%%
 The 't Hooft equation complements the light-front holographic Schrödinger equation in governing the longitudinal dynamics of quark-antiquark mesons. Together, they provide good predictions for the mass spectroscopy of the $\phi$-meson family without requiring additional parameter adjustments. Specifically, these calculations use the universal transverse confinement scale $\kappa=0.523$ GeV, the longitudinal confinement scale $g=0.109$ GeV~\cite{Ahmady:2022dfv}, and the strange quark mass $m_s=0.357$ GeV adopted in holographic light-front QCD (hLFQCD) along with the invariant mass ansatz (IMA) framework~\cite{Brodsky:2014yha}.

When combined with the color glass condensate dipole cross-section, the $\phi$-meson holographic light-front wavefunctions (LFWFs), incorporating the longitudinal mode from the 't Hooft equation, provide a good description of cross-section data for diffractive $\phi$-meson electroproduction at various energies. Using these resulting LFWFs, we have calculated key properties of the $\phi$-meson, including the decay constant, distribution amplitude, electromagnetic form factors (EMFFs), charge radius, magnetic moment, and quadrupole moment.
Interestingly, while the EMFFs predicted by our approach align well with those from hLFQCD-IMA, the two approaches yield different results for the distribution amplitudes. Additionally, our predictions for the vector and tensor decay constants are lower than the experimental measurement and various theoretical predictions; however, their ratio shows good agreement with values reported in the literature. Finally, the static properties, such as the charge radius and magnetic moment, are consistent with other theoretical results.

%%%%%%%%%%%%%%%%%
\section*{Acknowledgement}
%%%%%%%%%%%%%%%%%
We thank Ruben Sandapen and Mohammad Ahmady for fruitful discussions. C.M. is supported by new faculty start up funding by the Institute of Modern Physics, Chinese Academy of Sciences, Grant No.
E129952YR0. C.M. also thanks the Chinese Academy of
Sciences Presidents International Fellowship Initiative for the support via Grant No. 2021PM0023. S.K. is supported by Research Fund for
International Young Scientists, Grant No. 12250410251, from the National Natural Science Foundation of China
(NSFC), and China Postdoctoral Science Foundation
(CPSF), Grant No. E339951SR0.

\bibliography{ref.bib}

\end{document}